\documentclass[seceq,preprint]{ptptex}

\notypesetlogo  
\preprintnumber[3cm]{KYUSHU-HET-114}

\pubinfo{Vol. XXX, No.XX, XXX 20XX}

\title{
Strong CP Problem and the~Natural~Hierarchy of Yukawa~Couplings
}

\author{
Kenzo INOUE \footnote{E-mail: inoue@phys.kyushu-u.ac.jp}
and Naoki YAMATSU\footnote{E-mail: yamatsu@higgs.phys.kyushu-u.ac.jp}
}

\inst{
Department of Physics, Kyushu University, Fukuoka 812-8581, Japan
}

\recdate{XXXXX XX, 20XX
}

\abst{
A possible solution to the strong CP problem is presented
without using an axion.
The model is based on the framework of the supersymmetric
vectorlike theory with the spontaneous
 breakdown of the P-C-T-invariance.
It is shown that the characteristic structure of the Yukawa coupling
matrices 
that results from the spontaneous P-C-T-breaking plays the essential
role in naturally realizing $\theta=0$ at the tree level.
It is argued that $\theta=0$ will not be affected by 
the radiative corrections.
}

\begin{document}
\maketitle

\section{Introduction}

One of the prominent issues of the quantum chromo-dynamics
is the strong CP problem.\cite{ref:strong-CP}
The gauge principle based on the gauge group $SU_3\times SU_2\times U_1$ 
of the standard model or the minimal supersymmetric
standard model (MSSM) \cite{ref:MSSM,ref:MSSM-1,ref:MSSM-2,ref:MSSM-review}
does not forbid the appearance of the 
gluon $\theta$-term that violates the invariance under the
space inversion (P) and the time reversal (T):
\begin{equation}
{\cal L}_\theta=\theta \frac{g_c^2}{64\pi^2}\epsilon^{\mu\nu\lambda\sigma}
G^a_{\mu\nu}G^a_{\lambda\sigma} .
\label{theta-term}
\end{equation}
The upper limit of $\theta$ is severely constrained by some experiments.
The present limit is $|\theta|\lesssim 10^{-10}$.\cite{ref:PDG}
The promising candidate for the solution to the problem 
is an (invisible) axion,\cite{ref:axion,ref:invisible-axion}
which originates from the spontaneous 
breakdown of the Peccei-Quinn $U(1)_{\rm PQ}$ symmetry.\cite{ref:PQ_symmetry}
For the axion to solve this problem, the $U(1)_{\rm PQ}$ symmetry 
must be an anomalous symmetry that
suffers from the gauge anomaly of gluons.

On the other hand,
the recent progress of the
superstring theories suggests that
the various types of the superstring theories are
connected in the more fundamental theory. 
If the theory really admits
the chiral as well as the anti-chiral string theories,
it might not be unreasonable to imagine that
the original theory is the vectorlike theory.
The chiral nature of each string theory seems to be realized
through the spontaneous breakdown of the P-symmetry.
If this is the case, the $U(1)_{\rm PQ}$ symmetry, even if exists,
may not suffer from the gauge anomalies.
This gives the motivation to solve the strong CP problem
in an alternative way without using the axion.
So far, various attempts have been made 
for the spontaneous CP violation,\cite{ref:spontaneous-CP}
where the $\theta$-term is
sufficiently suppressed but the sizable 
CP violating phases in the CKM matrix \cite{ref:CKM}
are reserved.

We have examined, in the series of publications, 
\cite{ref:su11-1,ref:su11-2,ref:su11-3}
the vectorlike model\cite{ref:left-right_symmetry}
that realizes the MSSM as the low-energy effective theory 
through the spontaneous breakdown of the P-C-T-invariance.
In the recent publication,\cite{ref:su11-3}
we suggested that the model may
 solve the strong CP problem.
In this paper,
we will pursue the extensive analysis of this problem and 
clarify what is necessary to solve the problem.
We omit the lepton sector, 
which is essentially irrelevant to the subject.

\section{Setup of the model}

The model is based on the supersymmetric
vectorlike theory with the gauge group
$SU_3\times SU_2\times U_1\times SU(1,1)$, 
where the gauge group $SU(1,1)$ is a horizontal symmetry
\cite{ref:horizontal_symmetry,ref:Frog-Nielsen}
governing the generational structures of quarks and leptons.
The basic hypothesis is that the model is invariant under
the P, C, and T-transformations at the fundamental level.
Therefore, the original value of $\theta$ is $\theta_0=0$.
The model realizes the MSSM through the spontaneous breakdown of $SU(1,1)$.
The nonvanishing vacuum expectation values (VEVs) 
that break $SU(1,1)$ are shared by the some sets of the 
finite-dimensional nonunitary $SU(1,1)$
multiplets $\Psi$'s of a type
\begin{equation}
\Psi=\{\psi_{-S}, \psi_{-S+1}, \cdots, 
\psi_{S-1}, \psi_{S}\}.
\end{equation}
They are $SU_3\times SU_2\times U_1$ singlet
and their VEVs are  assumed to be roughly on the order of
$M\simeq 10^{16}$GeV to reproduce the successful MSSM \cite{ref:MSSM-success}
at the low-energy.
Although we are not yet able to give the explicit form
of their superpotential $W[\mbox{finite dim.}]$,
we make reasonable assumptions on their VEVs based on the intuitive
considerations.

The first assumption is that any of the multiplets 
$\Psi$'s takes its nonvanishing 
VEV at most at its single component $\langle\psi_M\rangle$
with some $SU(1,1)$ weight $M$.
Under the $U(1)_H$ transformation that is a subgroup of $SU(1,1)$,
$\psi_M$ transforms as
\begin{equation}
\psi_M\rightarrow e^{iM\varphi_H}\psi_M.
\end{equation}
The second assumption is
that any quantity $r_0$ related to the VEVs 
that has ``total weight 0'', such as 
\begin{equation}
r_0=\langle\psi_0\rangle,\ 
\frac{\langle\psi_M\rangle}{\langle\psi^\prime_M\rangle},\ 
\frac{\langle\psi_M\rangle\langle\psi^\prime_{N-M}\rangle}
{\langle\psi^{\prime\prime}_P\rangle
\langle\psi^{\prime\prime\prime}_{N-P}\rangle}
,\ \mbox{etc.},
\end{equation}
has a ``natural phase''
\begin{equation}
\frac{r_0}{|r_0|}=e^{i\pi p/q}
\label{natural-phase}
\end{equation}
with some set of integers $p$ and $q$, 
 since $W[\mbox{finite dim.}]$ does not 
contain any explicitly complex number.

The superpotential for the quark sector of the model is 
\begin{eqnarray}
W_{\mbox{quark}}&=&
(x_QQ_\alpha\bar Q_{-\alpha}+x_U\bar U_\beta U_{-\beta}
+x_D\bar D_\gamma D_{-\gamma})\Psi^F \nonumber\\
&&+(x^\prime_QQ_\alpha\bar Q_{-\alpha}+x^\prime_U\bar U_\beta U_{-\beta}
+x^\prime_D\bar D_\gamma D_{-\gamma})\Psi^{\prime F} .
\label{W-quark}
\end{eqnarray}
The multiplet $Q_\alpha$, for example, carries the $SU_3\times SU_2\times U_1$
quantum numbers of the quark doublet $q$ and belongs to the 
infinite-dimensional unitary representation of $SU(1,1)$ with
the positive 
lowest weight $\alpha$, and $\bar Q_{-\alpha}$ is its conjugate:
\begin{equation}
Q_\alpha=\{q_\alpha, q_{\alpha+1}, \cdots\},\hspace{1em}
\bar Q_{-\alpha}=\{\bar q_{-\alpha}, \bar q_{-\alpha-1}, \cdots\}.
\end{equation}
The multiplets $\Psi^F$ and $\Psi^{\prime F}$ 
belong to the finite-dimensional nonunitary representations
of $SU(1,1)$ with the common highest weight $S^F(\geq 3)$.
All coupling constants $x$'s and $x^\prime$'s in (\ref{W-quark})
are real numbers under the basic hypothesis
of the P-C-T-invariance.
The nonvanishing VEVs
\begin{equation}
\langle\Psi^F\rangle=\langle\psi^F_{-3}\rangle,\hspace{1em}
\langle\Psi^{\prime F}\rangle=\langle\psi^{\prime F}_0\rangle 
\end{equation}
in the superpotential (\ref{W-quark}) generate
three generations of the chiral quarks 
$q_m$, $\bar u_m$, and $\bar d_m$ \ $(m=0,1,2)$ that are embedded in
the infinite number of the 
components of $Q_\alpha$, $\bar U_\beta$, and $\bar D_\gamma$, respectively,
in the manner
\begin{equation}
q_{\alpha+i}=\sum_{m=0}^2q_mU^q_{mi}+[\cdots],
\hspace{1em}
\bar u_{\beta+i}=\sum_{m=0}^2\bar u_mU^u_{mi}+[\cdots],
\hspace{1em}
\bar d_{\gamma+i}=\sum_{m=0}^2\bar d_mU^d_{mi}+[\cdots],
\label{q-mixing}
\end{equation}
where $[\cdots]$ represents the superheavy massive modes.
The mixing coefficient $U^q_{mi}$, for example, is derived by the
requirement that $q_m$'s disappear from the mass operators
$x_QQ_\alpha\bar Q_{-\alpha}\langle\Psi^F\rangle+x^\prime_Q
Q_\alpha\bar
Q_{-\alpha}\langle\Psi^{\prime F}\rangle$.
This gives \cite{ref:su11-3}
\begin{equation}
U^q_{mi}=U^q_m\sum_{r=0}^\infty \delta_{i,m+3r}(-\epsilon_q)^rb^q_{mr}(\alpha),
\hspace{1em}m=0,1,2,
\label{U^q}
\end{equation}
with
\begin{equation}
\epsilon_q=\frac{x_Q^\prime}{x_Q}
\frac{\langle\psi_0^{\prime F}\rangle}{\langle\psi_{-3}^F\rangle} .
\end{equation}
The function $b^q_{mr}(\alpha)$ is a real function of the $SU(1,1)$
Clebsch-Gordan (C-G) coefficients depending on the weights of the multiplets.
We note that the mixing parameters
$\epsilon_q$, $\epsilon_u$, and $\epsilon_d$ are in general
complex numbers, but they have a common phase. 
Under the $U(1)_H$ transformation,
they transform as if they have a ``weight 3''. 
Since each component of $Q_\alpha$, $\bar U_\beta$, and $\bar D_\gamma$
transforms as
\begin{equation}
q_{\alpha+i}\rightarrow e^{i(\alpha+i)\varphi_H}q_{\alpha+i},\ \ 
\bar u_{\beta+i}\rightarrow e^{i(\beta+i)\varphi_H}\bar u_{\beta+i},\ \ 
\bar d_{\gamma+i}\rightarrow e^{i(\gamma+i)\varphi_H}\bar d_{\gamma+i},
\end{equation}
$q_m$, $\bar u_m$, and $\bar d_m$ $(m=0,1,2)$ transform under $U(1)_H$
as
\begin{equation}
q_m\rightarrow e^{i(\alpha+m)\varphi_H}q_m,\ \ 
\bar u_m\rightarrow e^{i(\beta+m)\varphi_H}\bar u_m,\ \ 
\bar d_m\rightarrow e^{i(\gamma+m)\varphi_H}\bar d_m.
\label{quark-tr}
\end{equation}
This means that the coefficients 
$U^q_m$, $U^u_m$, and $U^d_m$ do not receive any transformation.
We fix a phase convention of the chiral quarks
so that $U^q_m$, $U^u_m$, and $U^d_m$ are real and positive.

For the higgs sector, it has been shown that the model must have at
least double structure for the down-type higgs doublet $h'$.
\cite{ref:su11-3}
In this paper, we assume that both of the up-type and the down-type
higgses have the double structure, and introduce 
the $SU(1,1)$ multiplets
$H_{-\rho}$, $K_{-\rho-\Delta}$ ($\rho=\alpha+\beta$, $\rho+\Delta>0$)
for the up-type higgs doublet $h$,
and $H^\prime_{-\sigma}$, $K^\prime_{-\sigma-\Delta^\prime}$ 
($\sigma=\alpha+\gamma$, $\sigma+\Delta^\prime>0$)
for the down-type higgs doublet $h^\prime$, and 
all of their conjugates.
$\Delta$ and $\Delta^\prime$ are restricted to be an integer (Type-I) or 
a half-integer (Type-II).
The superpotential for the higgs sector is
\begin{eqnarray}
W_{\mbox{higgs}}&=&H\bar H\Psi+K\bar K\Phi+K\bar HX+H\bar K\Omega
\nonumber\\
&&+H^\prime\bar H^\prime\Psi^\prime+K^\prime\bar K^\prime\Phi^\prime
+K^\prime\bar H^\prime X^\prime+H^\prime\bar K^\prime\Omega^\prime ,
\label{Higgs}
\end{eqnarray}
where all real coupling constants have been
absorbed into the finite-dimensional nonunitary multiplets 
$\Psi\sim\Omega^\prime$.
To preserve the vectorlike nature of the model, $X$ and $\Omega$,
and also $X^\prime$ and $\Omega^\prime$ must be assigned to the common
$SU(1,1)$ representations.
It has been shown \cite{ref:su11-3}
 that there are $4\times 4$ cases for the pattern
of the VEVs $\langle\Psi\rangle\sim\langle\Omega^\prime\rangle$
that realize just
one set of chiral $h$ and $h^\prime$ in $H$, $K$ and 
$H^\prime$, $K^\prime$ in the manner
\begin{eqnarray}
&&h_{-\rho-i}=hU_i+[\cdots],\hspace{1.9em}
k_{-\rho-\Delta-i}=hV_i+[\cdots],\label{h-mixing}\\
&&h^\prime_{-\sigma-i}=h^\prime U^\prime_i+[\cdots],\hspace{1.5em}
k^\prime_{-\sigma-\Delta^\prime-i}=h^\prime V^\prime_i+[\cdots]
\label{h^prime-mixing} .
\end{eqnarray}
The mixing coefficients $U_i$, $V_i$, $U_i^\prime$, and $V_i^\prime$
have a general form
\begin{eqnarray}
&&U_i=U_0\epsilon^ib_i(\rho),\hspace{1.7em}
V_i=-r_{-\Delta}R_i(\rho)U_i,\label{UV}\\
&&U^\prime_i=U^\prime_0\epsilon^{\prime i}b^\prime_i(\sigma),\hspace{1.5em}
V^\prime_i=-r^\prime_{-\Delta^\prime}
R_i^\prime(\sigma)U^\prime_i,
\end{eqnarray}
where $b_i(\rho)$, $b^\prime_i(\sigma)$, $R_i(\rho)$, and 
$R^\prime_i(\sigma)$ are the real functions of the 
C-G coefficients, $\epsilon$ and $\epsilon^\prime$ have a
weight $-1$, and $r_{-\Delta}$ and $r^\prime_{-\Delta^\prime}$
have a weight $-\Delta$ and $-\Delta^\prime$, respectively. 
For example, a set of VEVs
\begin{eqnarray}
&&\langle\Psi\rangle=\langle\psi_1\rangle,\ \ \
\langle\Phi\rangle=\langle\phi_0\rangle,\ \ \
\langle X\rangle=\langle\kappa_\Delta\rangle,\ \ \
\langle\Omega\rangle=\langle\omega_{-\Delta}\rangle ,
\label{VEVs-for-h}
\\
&&\langle\Psi^\prime\rangle=\langle\psi^\prime_0\rangle,\ \ 
\langle\Phi^\prime\rangle=\langle\phi^\prime_0\rangle,\ \ 
\langle X^\prime\rangle=\langle\kappa^\prime_{\Delta^\prime}\rangle,\ \ \
\langle\Omega^\prime\rangle=\langle\omega^\prime_{1-\Delta^\prime}\rangle 
\label{VEVs-for-h'}
\end{eqnarray}
gives 
\begin{equation}
\epsilon=\frac{\langle\chi_\Delta\rangle\langle\omega_{-\Delta}\rangle}
{\langle\psi_1\rangle\langle\phi_0\rangle},~~~
\epsilon^\prime=
\frac{\langle\psi^\prime_0\rangle\langle\phi^\prime_0\rangle}
{\langle\chi^\prime_{\Delta^\prime}
\rangle\langle\omega^\prime_{1-\Delta^\prime}\rangle},~~~
r_{-\Delta}=\frac{\langle\omega_{-\Delta}\rangle}{\langle\phi_0\rangle},~~~
r^\prime_{-\Delta^\prime}
=\frac{\langle\psi^\prime_0\rangle}
{\langle\chi^\prime_{\Delta^\prime}\rangle}.
\end{equation}
Notice that $h$ and $h^\prime$ transform under the $U(1)_H$
transformation as
\begin{equation}
h\rightarrow e^{-i\rho\varphi_H}h,\ \ h^\prime\rightarrow
e^{-i\sigma\varphi_H}h^\prime,
\label{higgs-tr}
\end{equation}
and $U_0$ and $U_0^\prime$ do not receive any transformation.
For the phase convention of
$h$ and $h^\prime$, 
we take $U_0$ and $U_0^\prime$ to be real and positive.

To reproduce the MSSM in the low-energy,
we need the additional sector that generates the 
$\mu$-term $\mu hh^\prime$.
It should be noticed that
the coupling of the form $HH^\prime\Psi$ is forbidden 
by the $SU(1,1)$ symmetry, since both $H$ and $H^\prime$
have negative weights.
This fact explains why $\mu$ does not take
a huge mass scale $M\simeq 10^{16}$GeV. 
Since the order of $\mu$ should be the supersymmetry breaking scale
$m_{\rm SUSY}\simeq 10^{2\sim 3}$GeV, what we have to do in the
supersymmetric limit is to generate the effective 
superpotential $W_\mu$ that is
sensitive to the supersymmetry breaking.
This means that $W_\mu$ contains the massless particles.
The simplest procedure for this sector is
to introduce the superpotential
that contains the $SU_3\times SU_2\times U_1$ singlets
$R_{(\rho+\sigma)/2}$, $R^\prime_{(\rho+\sigma)/2+1}$, 
$S_{\rho+\sigma}$, and their conjugates:
\begin{equation}
W_{\rm M}=(R\bar R^\prime+R^\prime\bar R)\Psi^M+S\bar S\Psi^S
+\tilde y_1\hspace{-.2em}
\left(HH^\prime S+\bar H\bar H^\prime\bar S
\right)
+\tilde y_2\hspace{-.2em}\left(RR\bar S+\bar R\bar RS
\right) ,
\label{MU}
\end{equation}
where the coupling constants $\tilde y_1,\tilde y_2\simeq O(1)$ 
are real numbers.
Notice that the superpotentials (\ref{Higgs}) and (\ref{MU}) respect
the Peccei-Quinn $U(1)_{\rm PQ}$
symmetry with the charges given by
\begin{equation}
\begin{array}{c|ccccccc|c} & H & K & H^\prime & K^\prime 
& R & R^\prime & S & \Psi\mbox{'s}\\
\hline
Q_{\rm PQ} & -1 & -1 & -1 & -1 & 1 
& 1 & 2 & 0
\end{array}\ .
\label{PQ-1}
\end{equation}
All finite-dimensional multiplets $\Psi$'s are assigned to be 
$Q_{\rm PQ}=0$.
Thus, each of the conjugate multiplets has opposite $Q_{\rm PQ}$
charge to that of the corresponding multiplets in (\ref{PQ-1}).  
This means that the $U(1)_{\rm PQ}$ symmetry is a vector symmetry, which
is free from any gauge anomalies in the framework of the vectorlike
gauge theory. 
Through the VEV $\langle\psi_0^M\rangle$ in (\ref{MU}),
the first component $r$ of $R$ and $\bar r$ of $\bar R$
become massless.
The VEV $\langle\psi_0^S\rangle$
gives huge masses to all components of $S$ and $\bar S$.
Notice that only the first component $\bar s$ of $\bar S$
couples to $rr$ and the first component $s$ of $S$
couples to $\bar r\bar r$ and $hh^\prime$.
Therefore, the relevant part of the superpotential is
\begin{equation}
\widetilde{W}_{\rm M}=M_ss\bar s+y_1U_0U_0^\prime
hh^\prime s +y_2(rr\bar s+\bar r\bar rs),
\label{tilde-W_M}
\end{equation}
where 
\begin{equation}
M_s=D^S_0\langle\psi_0^S\rangle,\hspace{1em}
y_1=\tilde y_1C^H_{0,0},\hspace{1em}
y_2=\tilde y_2C^R_{0,0},
\label{M_s}
\end{equation}
with the real and positive C-G coefficients $D^S_0$, $C^H_{0,0}$,
and $C^R_{0,0}$.
The integration of the superheavy $s$ and $\bar s$ leads to
the effective superpotential
\begin{equation}
W_\mu=-\frac{y_2}{M_s}rr(y_1U_0U_0^\prime hh^\prime+y_2\bar r\bar r).
\label{W_mu}
\end{equation}

Now, let us make the phenomenologically desirable requirements
on the characteristics of the supersymmetry breaking.
We assume that the supersymmetry breaking occurs in the hidden sector
that is singlet under $SU_3\times SU_2\times U_1\times SU(1,1)$,
and does not give the CP-violating phases to the observable
sector that consists of $F$'s, $\bar F$'s, $\Psi$'s, 
and the relevant gauge multiplets.
Thus, the gauginos have real masses at the energy scale 
$E\simeq 10^{16}$GeV.
For the K\"ahler potential, we assume the form
\begin{equation}
K=\sum_Ff_F(z_i,z_i^*)\left(F^\dagger F
+\bar F^\dagger \bar F \right)
+\sum_Ag_A(z_i,z_i^*)K_A(\Psi\mbox{'s},\Psi^\dagger\mbox{'s})
+k_H(z_i,z_i^*)
\label{Kaeler}
\end{equation} 
up to gauge couplings, where $f_F$, $g_A$,
and $k_H$ are real functions
of the hidden sector fields $z_i$.
Even if we assume (\ref{Kaeler}),
we must take into account the tree diagrams connected by the superheavy
particles.
For example, the superheavy $\bar Q$ inevitably induces the term
\begin{equation}
\Delta K\propto Q^\dagger
(x_Q{\Psi^F}^\dagger+x'_Q{\Psi^{\prime F}}^\dagger)
(x_Q\Psi^F+x'_Q\Psi^{\prime F})Q ,
\label{K-add}
\end{equation}
because $\Psi^F$'s couple to $Q$ and $\bar Q$ in the superpotential
(\ref{W-quark}).
The chiral modes $q_m$'s, however, are realized so that they decouple
from the VEVs of $\Psi^F$'s.
This means that $q_m$'s disappear in (\ref{K-add}) 
when $\Psi^F$'s in (\ref{K-add}) are replaced by their VEVs.
Therefore, as far as the chiral modes are concerned, 
the  K\"ahler potential (\ref{Kaeler}) will be appropriate.
Thus, we expect that
all chiral modes in the matter multiplets $F$'s and $\bar F$'s 
have the generation-independent
soft masses $m^2_F(=m^2_{\bar F})$ at $E\simeq 10^{16}$GeV.
The K\"ahler potential (\ref{Kaeler}) gives the further
consequence on the A-terms.
Since the hidden sector fields are $SU(1,1)$ singlets,
A-terms should respect the $SU(1,1)$ symmetry.
Therefore, each of the A-terms must be exactly proportional,
at $E\simeq 10^{16}$GeV,
to the corresponding term in the superpotential.

Under this setup, the supersymmetry breaking terms 
in the bosonic potential that contain $r$ and $\bar r$ are given by
\begin{equation}
V_{\rm s.b.}(r,\bar r)=m_r^2(|r|^2+|\bar r|^2)
-\left(\frac{y_2m_0}{M_s}rr(y_1A_1U_0U_0^\prime hh^\prime+y_2A_2\bar r\bar r)
+\mbox{h.c.}\right) ,
\label{V(r)}
\end{equation}
where $m_r,m_0\ (\simeq m_{\rm SUSY})$ are real and positive, 
and $A_1,A_2\ (\simeq O(1))$ are real values.
Since $r$ and $\bar r$ are the $SU_3\times SU_2\times U_1$ singlets,
their tree-level potential takes a form
\begin{equation}
V(r,\bar r)=4\frac{y_2^4}{|M_s|^2}|r\bar r|^2
(|r|^2+|\bar r|^2)+m_r^2(|r|^2+|\bar r|^2)
-\left(y_2^2A_2\frac{m_0}{M_s}rr\bar r\bar r+\mbox{h.c.}\right) .
\label{V(r)-full}
\end{equation}
This potential has three local minima when 
$|A_2|>2\sqrt{3}\hspace{.1em}m_r/m_0$.
One is a trivial minimum at the origin.
The other two minima are degenerate and sit on the circles
specified by
\begin{equation}
r\eta^*=\pm\bar r\eta\varepsilon
=\frac{\sqrt{|A_2|+\sqrt{|A_2|^2-12(m_r/m_0)^2}}}{2\sqrt{3}\ |y_2|}
\sqrt{m_0|M_s|}\equiv w_0,
\label{VEV-r}
\end{equation}
where
\begin{equation}
|\eta|=1,
\hspace{1em}|\varepsilon|=1,\hspace{1em}
\arg[\varepsilon]=\frac{1}{2}\arg\left[\frac{A_2}{M_s}\right],
\label{eta}
\end{equation}
and $\eta$ is a unimodulus constant representing the 
spontaneous breakdown of the $U(1)_{\rm PQ}$ symmetry.
Its phase freedom represents the associated Nambu-Goldstone (N-G) boson.

When $|A_2|>4\hspace{.1em}m_r/m_0$, 
these minima become deeper than the origin.
The VEVs (\ref{VEV-r}) generate the $\mu$-term $\mu hh^\prime$ 
in (\ref{W_mu}) 
and the $B$-term $\mu Bhh^\prime$ in (\ref{V(r)}) with the
desirable magnitudes:
\begin{equation}
\mu=-2U_0U_0^\prime\frac{y_1y_2w_0^2}{M_s},\hspace{2em}
B=m_0A_1,
\label{mu,B}
\end{equation}
where we have taken the phase convention $\eta=1$.
Notice that $B$ takes a real value.
Expressing $r$ and $\bar r$ as
\begin{equation}
r=w_0+\tilde r,\hspace{1.5em}
\bar r=\varepsilon^*(w_0+\tilde{\bar r})
\end{equation}
and substituting them for (\ref{W_mu})
and (\ref{V(r)}), we find that $\tilde r$ and $\tilde{\bar r}$
have the mass terms on the order of $m_{\rm SUSY}$ and
the coupling $\tilde rhh^\prime$
is suppressed by the factor $\sqrt{m_{\rm SUSY}/M}\simeq 10^{-7}$.
Thus, the effective theory below the energy scale 
$\sqrt{m_{\rm SUSY}M}$ is described by the MSSM
and the almost decoupled neutral $\tilde r$ and $\tilde{\bar r}$.

\section{Natural hierarchy of the Yukawa couplings}

Let us proceed to the Yukawa couplings of the higgses to the quarks.
Owing to the weight constraint of the $SU(1,1)$ symmetry,
the superpotential $W_{\rm Y}$ that describes the Yukawa couplings
takes the different form for the integer $\Delta$, $\Delta'$ 
(Type-I) and for the half-integer $\Delta$, $\Delta'$ (Type-II):
\cite{ref:su11-3}
\begin{eqnarray}
&&\mbox{Type-I}:\hspace{1.0em}
W_{\rm Y}=y_U\bar UQH+y_U^\Delta \bar UQK
+y_D\bar DQH^\prime+y_D^{\Delta^\prime} \bar DQK^\prime 
+[\mbox{mirror couplings}],\nonumber\\
\label{Yukawa-I}\\
&&\mbox{Type-II}:\hspace{0.6em}
W_{\rm Y}=y_U\bar UQH+y_D\bar DQH^\prime
+[\mbox{mirror couplings}]
\label{Yukawa-II}.
\end{eqnarray}
Notice that these superpotentials also preserve the $U(1)_{\rm PQ}$
symmetry with the charges
\begin{equation}
\begin{array}{c|ccc} & Q & \bar U & \bar D \\
\hline
Q_{\rm PQ} & 1/2+\delta_B & 1/2-\delta_B & 1/2-\delta_B 
\end{array},
\label{PQ-2}
\end{equation}
in addition to (\ref{PQ-1}),
where $\delta_B$ represents a contamination of the baryon number
charge. 
We note that both of the $U(1)_{\rm PQ}$ symmetry and the baryon number
$U(1)_B$ symmetry are free from gauge anomalies because they are vector
symmetries. 

Substituting the expressions (\ref{q-mixing}), (\ref{h-mixing}), and
(\ref{h^prime-mixing}) for $W_{\rm Y}$
and extracting the massless modes,
we obtain the effective superpotential
for the Yukawa couplings of
the higgses $h$, $h^\prime$ to
the quarks $q_m$, $\bar u_m$, $\bar d_m$
($m=0,1,2$):
\begin{equation}
W_{\rm Yukawa}=\sum_{m,n=0}^2\left(y_u^{mn}\bar u_m q_n h
+y_d^{mn}\bar d_m q_n h^\prime\right) .
\label{yukawa}
\end{equation}
For the contraction of two $SU_2$ indices, we adopt
a convention $qh\equiv \varepsilon_{ij}q^ih^j$
with $\varepsilon_{12}=1$, $\varepsilon_{21}=-1$.

The coupling matrix $y_u^{mn}$ for the Type-I scheme is expressed as
\begin{equation}
y_u^{mn}=y_U\sum_{i,j=0}^\infty C^{\beta,\alpha}_{i,j}(0)
U^u_{mi}U^q_{nj}U_{i+j}
+y_U^\Delta\sum_{i,j=0}^\infty C^{\beta,\alpha}_{i,j}(\Delta)
U^u_{mi}U^q_{nj}V_{i+j-\Delta},
\end{equation}
where $C^{\beta,\alpha}_{i,j}(0)$ and $C^{\beta,\alpha}_{i,j}(\Delta)$
are the real C-G coefficients satisfying the symmetry relation
\cite{ref:su11-3}
\begin{equation}
C^{\beta,\alpha}_{i,j}(\Delta)=(-1)^\Delta C^{\alpha,\beta}_{j,i}(\Delta).
\label{C-sym}
\end{equation} 
The expression of the mixing coefficients (\ref{U^q})
and (\ref{UV}) then gives
\begin{equation}
y_u^{mn}=y_UU^u_mU^q_nU_0\epsilon^{m+n}Y_u^{mn}
\label{y-u}
\end{equation}
with
\begin{equation}
Y_u^{mn}=A_u^{mn}-r_UB_u^{mn} ,\hspace{2em}
r_U=\frac{y_U^\Delta}{y_U}\epsilon^{-\Delta}r_{-\Delta},
\label{Y^u}
\end{equation}
where
\begin{eqnarray}
&&A_u^{mn}=\sum_{r,s=0}^\infty(-\epsilon_u\epsilon^3)^r
(-\epsilon_q\epsilon^3)^sC^{\beta,\alpha}_{m+3r,n+3s}(0)
b^u_{mr}(\beta)b^q_{ns}(\alpha)b_{m+n+3(r+s)}(\rho) ,
\label{A-u}\\
&&B_u^{mn}=\sum_{r,s=0}^\infty\theta_{m+n+3(r+s),\Delta}
(-\epsilon_u\epsilon^3)^r
(-\epsilon_q\epsilon^3)^sC^{\beta,\alpha}_{m+3r,n+3s}(\Delta)
\nonumber\\
&&\hspace{6em}\times 
b^u_{mr}(\beta)b^q_{ns}(\alpha)b_{m+n+3(r+s)-\Delta}(\rho)
R_{m+n+3(r+s)-\Delta}(\rho) ,\label{B-u}
\end{eqnarray}
and $\theta_{m+n+3(r+s),\Delta}=1$ for $m+n+3(r+s)\ge\Delta$ and $0$
for $m+n+3(r+s)<\Delta$.

The coupling matrix $y_d^{mn}$ takes a form
\begin{equation}
y_d^{mn}=y_DU^d_mU^q_nU^\prime_0\epsilon^{\prime m+n}Y_d^{mn}
\label{y-d}
\end{equation}
with
\begin{equation}
Y_d^{mn}=A_d^{mn}-r_DB_d^{mn} ,\hspace{2em}
r_D=\frac{y_D^{\Delta^\prime}}{y_D}
\epsilon^{\prime-\Delta^\prime}r^\prime_{-\Delta^\prime},
\label{Y^d}
\end{equation}
where
\begin{eqnarray}
&&A_d^{mn}=\sum_{r,s=0}^\infty(-\epsilon_d\epsilon^{\prime 3})^r
(-\epsilon_q\epsilon^{\prime 3})^sC^{\gamma,\alpha}_{m+3r,n+3s}(0)
b^d_{mr}(\gamma)b^q_{ns}(\alpha)b^\prime_{m+n+3(r+s)}(\sigma) ,
\label{A-d}\\
&&B_d^{mn}=\sum_{r,s=0}^\infty\theta_{m+n+3(r+s),\Delta^\prime}
(-\epsilon_d\epsilon^{\prime 3})^r
(-\epsilon_q\epsilon^{\prime 3})^sC^{\gamma,\alpha}_{m+3r,n+3s}(\Delta^\prime)
\nonumber\\
&&\hspace{6em}\times 
b^d_{mr}(\gamma)b^q_{ns}(\alpha)b^\prime_{m+n+3(r+s)-\Delta^\prime}(\sigma)
R_{m+n+3(r+s)-\Delta^\prime}(\sigma) .\label{B-d}
\end{eqnarray}
Notice that all of $r_U$, $r_D$, $\epsilon_u\epsilon^3$,
$\epsilon_q\epsilon^3$, $\epsilon_d\epsilon^{\prime 3}$,
and $\epsilon_q\epsilon^{\prime 3}$ have a weight $0$.

The coupling matrices $y_u^{mn}$ and $y_d^{mn}$ for the Type-II
scheme are obtained 
by simply setting $r_U=r_D=0$ in (\ref{Y^u}) and (\ref{Y^d}).

The remarkable aspect of the model is that the 
coupling matrices $y_u^{mn}$
and $y_d^{mn}$ have the definite transformation property
under the $U(1)_H$ transformation:
\begin{equation}
y_u^{mn}\rightarrow e^{-i(m+n)\varphi_H}y_u^{mn},\hspace{2em}
y_d^{mn}\rightarrow e^{-i(m+n)\varphi_H}y_d^{mn} .
\label{y-H-tr}
\end{equation}
Although the Yukawa couplings (\ref{yukawa}) are only a subset of the
$SU(1,1)$ invariant Yukawa couplings $W_{\rm Y}$ given by (\ref{Yukawa-I})
or (\ref{Yukawa-II}),
this transformation property, supplemented by (\ref{quark-tr})
and (\ref{higgs-tr}), assures  (\ref{yukawa}) to be invariant
under the $U(1)_H$ transformation.

\section{Toy model analysis}

Before discussing the details of the CP problem of the model,
we examine a toy model analysis that will give indispensable
information for the later considerations.
The toy model is based on the $SU_3\times SU(1,1)$ gauge group.

The superpotential of the model is 
\begin{eqnarray}
W&=&
(x_PP_\alpha\bar P_{-\alpha}+x_Q\bar Q_\beta Q_{-\beta}
)\Psi^F
+x_HH_{-\rho}\bar H_\rho(\Psi+\Psi')\nonumber\\
&&+y(P_\alpha\bar Q_\beta H_{-\rho}+\bar P_{-\alpha}Q_{-\beta}\bar H_{\rho})
+W[\mbox{finite dim.}] .
\label{W-toy}
\end{eqnarray}
The multiplets $P_\alpha$ and $Q_{-\beta}$ are the $SU_3$ triplets,
and $H_{-\rho}$ is the singlet:
\begin{equation}
P_\alpha=\{p_\alpha, p_{\alpha+1}, \cdots\},~
Q_{-\beta}=\{q_{-\beta}, q_{-\beta-1}, \cdots\},~
H_{-\rho}=\{h_{-\rho}, h_{-\rho-1}, \cdots\},~~
\rho=\alpha+\beta .
\end{equation}
The multiplets $\bar P_{-\alpha}$, $\bar Q_\beta$, and 
$\bar H_{\rho}$ are the conjugate
representations.
The multiplets $\Psi^F$, $\Psi$, and $\Psi'$ 
have the highest weights $S^F$, $S$, and $S'(=S)$, respectively.
The coupling constants $x_P$, $x_Q$, and $y$ in (\ref{W-toy})
are real values.

The superpotential (\ref{W-toy}) respects the global symmetry 
$U(1)_{\rm PQ}$ with the charges given by
\begin{equation}
\begin{array}{c|c|c|c|c} & P_\alpha & Q_{-\beta} & H_{-\rho} 
& \Psi\mbox{'s}\\
\hline
Q_{\rm PQ} & 1/2 & -1/2 & -1 & 0 
\end{array}
\end{equation}
in addition to the baryon number $U(1)_B$ symmetry.
Both of the $U(1)_{\rm PQ}$ and $U(1)_B$ symmetries may not suffer from
the $SU_3$ gauge anomaly because the model is purely vectorlike,
which is anomaly-free at any scale.
So, if superheavy modes decouple from the low-energy and 
the other modes generate the anomaly of the $U(1)_{\rm PQ}$ and
$U(1)_B$ symmetries,
then the anomaly matching condition\cite{ref:Anomaly-match} 
requires introducing the Wess-Zumino-Witten (WZW) terms\cite{ref:WZW}
in order to realize anomaly-free.
However, in the following analysis, we deal with all modes including
superheavy modes, so we need not consider the WZW terms.

The nonvanishing VEV
\begin{equation}
\langle\Psi^F\rangle=\langle\psi^F_{-g}\rangle,\hspace{1em}
\label{VEV-F}
\end{equation}
in the superpotential (\ref{W-toy}) generates
$g$ generations of the chiral quarks 
\begin{equation}
p_m\equiv p_{\alpha+m},\hspace{1em}
\bar q_m\equiv \bar q_{\beta+m},\hspace{1.5em}(m=0,1,\cdots,g-1).
\end{equation}
Other components
\begin{equation}
p_{I+g}\equiv p_{\alpha+I+g},~~
\bar q_{I+g}\equiv \bar q_{\beta+I+g},~~
\bar p_I\equiv \bar p_{-\alpha-I},~~
q_I\equiv q_{-\beta-I}~~~(I=0,1,2,\cdots)
\end{equation}
become superheavy.
The VEVs
\begin{equation}
\langle\Psi\rangle=\langle\psi_1\rangle,\hspace{1em} 
\langle\Psi'\rangle=\langle\psi'_0\rangle 
\label{VEVs}
\end{equation}
generate one chiral higgs $h$ in $H_{-\rho}$ in the form
\begin{equation}
h_{-\rho-i}=U_ih+[\mbox{massive modes}].
\end{equation}
The mixing coefficient $U_i$ takes a form
\begin{equation}
U_i=U_0\epsilon^ib_i(\rho),
\hspace{1em}
\epsilon=\frac{\langle\psi'_0\rangle}{\langle\psi_1\rangle},
\label{U-toy}
\end{equation}
where $b_i(\rho)$ is a real function of the C-G
coefficients.

The effective theory below the energy scale 
$E\simeq 10^{16}$GeV is described by the superpotential
\begin{equation}
W_{\rm eff}=\sum_{m,n=0}^{g-1}y\hspace{.1em}C^F_{m,n}U_{m+n}\bar q_mp_nh,
\end{equation}
where $C^F_{m,n}$ is a real C-G coefficient.
The supersymmetry breaking gives the breaking potential
\begin{equation}
V_{\rm s.b.}=
\sum_{m=0}^{g-1}(m_p^2|p_m|^2+m_q^2|\bar q_m|^2)+m_h^2|h|^2
+\left(\sum_{m,n=0}^{g-1}Am_0y\hspace{.1em}C^F_{m,n}U_{m+n}\bar q_mp_nh
+\mbox{h.c.}\right).
\end{equation}
The radiative corrections due to the Yukawa couplings
pull down $m_h^2$ to a negative value,\cite{ref:MSSM-1}
and $h$ eventually acquires
the nonvanishing VEV $\langle h\rangle$. 
This VEV brings the spontaneous breakdown of the $U(1)_{\rm PQ}$
symmetry, and the N-G boson $G^0$ appears.

Let us parametrize all VEVs in the form
\begin{equation}
\langle\psi^F_{-g}\rangle=e^{-ig\varphi_H}V^F,
\hspace{1em}
\langle\psi_1\rangle=e^{i\varphi_H}V,
\hspace{1em}
\langle\psi'_0\rangle=V',
\hspace{1em}
\langle h\rangle=e^{-i\rho\varphi_H}e^{-i\varphi_{\rm PQ}}v,
\label{VEVs-phi}
\end{equation}
where $\varphi_H$ represents the flat direction due to the $SU(1,1)$
symmetry, and $\varphi_{\rm PQ}$ represents the $U(1)_{\rm PQ}$ symmetry.
Due to the $SU(1,1)$ gauge invariance,
the total potential should be flat with respect to $\varphi_H$.
The spontaneous breakdown of the $SU(1,1)$ symmetry will take
some value of $\varphi_H$, but any physically observable
quantity should be independent of $\varphi_H$.

To evaluate $\theta$, let us write the tree level mass terms of 
the quarks:
\begin{eqnarray}
&&\sum_{m,n=0}^{g-1}y\hspace{.1em}C_{m,n}^FU_{m+n}\bar q_mp_n\langle h\rangle
+\sum_{I=0}^\infty x_P\hspace{.1em}D^P_Ip_{I+g}\bar p_I
\langle\psi^F_{-g}\rangle 
+\sum_{I=0}^\infty x_Q\hspace{.1em}D^Q_I\bar q_{I+g}q_I
\langle\psi^F_{-g}\rangle\nonumber\\
&&+\sum_{I=0}^\infty\sum_{n=0}^{g-1}y\hspace{.1em}
C_{I+g,n}^FU_{I+g+n}\bar q_{I+g}p_n\langle h\rangle
+\sum_{m=0}^{g-1}\sum_{I=0}^\infty y\hspace{.1em}
C_{m,I+g}^FU_{m+I+g}\bar q_mp_{I+g}\langle h\rangle\nonumber\\
&&+\sum_{I=0}^\infty\sum_{K=0}^{\infty}y\hspace{.1em}
C_{I+g,K+g}^FU_{I+K+2g}\bar q_{I+g}p_{K+g}\langle h\rangle\nonumber\\
&&\equiv \bar{\cal Q}{\cal M}_{\rm tree}{\cal Q},
\label{toy-mass-term}
\end{eqnarray}
where $D^P_I$ and $D^Q_I$ are real C-G coefficients and
\begin{eqnarray}
&&\bar{\cal Q}=\{\bar q_m|\bar q_{I+g}|\bar p_J\},~
{\cal Q}=\{p_n|p_{K+g}|q_L\},~~\nonumber\\
&&\hspace{5em}(m,n=0,1,\cdots,g-1\ ;\ I,J,K,L=0,1,\cdots),\\
&&{\cal M}_{\rm tree}
=\left(\begin{array}{c|c|c}y\hspace{.1em}C_{m,n}^FU_{m+n}
\langle h\rangle & y\hspace{.1em}C_{m,K+g}^FU_{m+K+g}
\langle h\rangle & 0 \\
\hline
y\hspace{.1em}C_{I+g,n}^FU_{I+g+n}\langle h\rangle & 
y\hspace{.1em}C_{I+g,K+g}^FU_{I+K+2g}\langle h\rangle
& x_QD_I^Q\langle\psi^F_{-g}\rangle \delta_{IL}\\
\hline
0 & x_PD^P_J\langle\psi^F_{-g}\rangle\delta_{JK}  & 0
\end{array}
\right).\nonumber\\
\label{toy-mass-matrix}
\end{eqnarray}
The diagonalization of this mass matrix will generate the
$\theta$-term (\ref{theta-term})
with $\theta$ presented by
\begin{equation}
\theta=\arg\det[{\cal M}_{\rm tree}].
\end{equation}
Since ${\cal M}_{\rm tree}$ is an infinite-dimensional matrix,
we do not have a reliable method for the calculation.
So, in the following considerations,
we take a heuristic approach based on the two alternative assumptions.

\subsection*{\it Naive evaluation}

First, we simply write
\begin{eqnarray}
\det[{\cal M}_{\rm tree}]&=&
-\det[y\hspace{.1em}C_{m,n}^FU_{m+n}\langle h\rangle]\times
\det\left[(x_PD^P_J\langle\psi^F_{-g}\rangle\delta_{JK})\times
(x_QD_K^Q\langle\psi^F_{-g}\rangle \delta_{KL})\right]\nonumber\\
&=&-\det[y\hspace{.1em}C_{m,n}^FU_{m+n}\langle h\rangle]\times
\det\left[x_Px_QD^P_JD_J^Q\langle\psi^F_{-g}\rangle
\langle\psi^F_{-g}\rangle \delta_{JK}\right]
\label{det[naive]}
\end{eqnarray}
following a formula for a finite-dimensional matrix.
Then, we have
\begin{equation}
\theta=\pi+
\arg\det[y\hspace{.1em}C_{m,n}^FU_{m+n}\langle h\rangle]
+\arg\det\left[x_Px_QD^P_JD_J^Q\langle\psi^F_{-g}\rangle
\langle\psi^F_{-g}\rangle \delta_{JK}\right].
\label{theta-1}
\end{equation}
Substituting the expressions (\ref{U-toy}) and (\ref{VEVs-phi}),
we find 
\begin{eqnarray}
&&\arg\det[y\hspace{.1em}C_{m,n}^FU_{m+n}\langle h\rangle]
\label{det-varphi_PQ}
\nonumber\\
&&\hspace{1em}=g(g-1)(-\varphi_H+\arg[V'/V])-g\rho\varphi_H
-g\varphi_{\rm PQ}+g\arg[v]\nonumber\\
&&\hspace{1em}+\arg\det[y\hspace{.1em}C_{m,n}^FU_0b_{m+n}(\rho)],\ \ \ \ \\
&&\arg\det\left[x_Px_QD^P_JD_J^Q\langle\psi^F_{-g}\rangle
\langle\psi^F_{-g}\rangle \delta_{JK}\right]\nonumber\\
&&\hspace{1em}=\arg\det[x_Px_Qe^{-i2g\varphi_H}(V^F)^2\delta_{JK}]
+\arg\det\left[D^P_JD_J^Q
\delta_{JK}\right].
\label{arg-PQ}
\end{eqnarray}
Although the C-G coefficients
$D_J^P$ and $D_J^Q$ have an alternating sign $(-1)^J$,
the product $D_J^PD_J^Q$ is real and positive.
So, the second term in (\ref{arg-PQ}) vanishes.
Let us introduce a numerical constant $c$ by
\begin{equation}
\arg\det[x_Px_Qe^{-i2g\varphi_H}(V^F)^2\delta_{JK}]
=c\left(-2g\varphi_H+2\arg[V^F]+\arg[x_Px_Q]\right).
\end{equation}
Formally, $c$ is a divergent quantity since we have
infinite numbers of the superheavy quarks:
\begin{equation}
c=1+1+1+\cdots.
\end{equation}
However, we do not mind it.
Summing up all terms in (\ref{theta-1}),
we obtain
\begin{eqnarray}
&&\theta=\pi-g\left(g-1+\rho+2c\right)
\varphi_H
-g\varphi_{\rm PQ}+g(g-1)\arg[V'/V]+g\arg[v]\nonumber\\
&&\hspace{2em}+c\left(2\arg[V^F]+\arg[x_Px_Q]\right)
+\arg\det[y\hspace{.1em}C_{m,n}^FU_0b_{m+n}(\rho)].
\label{theta-toy-tree}
\end{eqnarray}

Let us next consider the effect of the radiative corrections 
induced by the low-energy effective theory
to the quark mass matrix ${\cal M}_{\rm tree}$.
The wave function renormalization 
is irrelevant to $\theta$.\cite{ref:wave-func-to-theta}
This is because, when we perform the wave function renormalization,
we must change the path integral measure, and both of the effects sum 
up to vanish in $\theta$.
Therefore, what we must consider is a vertex correction to 
${\cal M}_{\rm tree}$.
Owing to the non-renormalization theorem,\cite{ref:Non-Renom.}
the origin of the correction is limited to the supersymmetry breaking.
This means that the corrections to the masses of 
the superheavy quarks are
suppressed by $m_{\rm SUSY}/M$.
Thus, the dominant effect is to the mass matrix
${\cal M}^{mn}$ of the $g$ generations of the chiral quarks.
The quark and/or squark loops, however,
do not give phases to ${\cal M}^{mn}$ 
since the loops contain the coupling matrix
$Y^{mn}\equiv C_{m,n}^FU_{m+n}$ in the real form
\begin{equation}
{\rm tr}[YY^\dagger Y\cdots Y^\dagger YY^\dagger].
\end{equation}
The (s)quark line connecting the external quarks contains $Y$ in the form
\begin{equation}
\left(YY^\dagger Y\cdots Y^\dagger Y\right)^{mn}.
\end{equation}
This array of matrices preserves the phase structure $U_{m+n}$ 
of the single $Y^{mn}$.
The number of the VEV $\langle h\rangle$ attached to the diagram
is larger than that of $\langle h\rangle^*$ just by 1.
Thus, the vertex correction does not 
disturb the phase structure of ${\cal M}^{mn}_{\rm tree}$.
The effect is limited to the modification of the last term in 
(\ref{theta-toy-tree}).
These arguments also apply to the superheavy quark masses.
The gluino mass is also real.
Therefore, (\ref{theta-toy-tree}) holds even after the 
radiative corrections of the low-energy physics
are fully included
as long as the determinant in the last term does not change its sign.

It is known that the $SU_3$ instantons give the vacuum energy
proportional to $(1-\cos\theta)$.\cite{ref:strong-CP}
Therefore, the $SU(1,1)$ invariance requires 
the coefficient of $\varphi_H$ in (\ref{theta-toy-tree}) to vanish.
Thus, we have, for the consistent value of $c$,
\begin{equation}
c=-\frac{g-1}{2}-\frac{\rho}{2},\hspace{2em}(\rho=\alpha+\beta).
\label{c-toy}
\end{equation}
The expression (\ref{theta-toy-tree}) gives further restriction
on the value of $c$.
Although we assume that the coupling constants $x_P$ and $x_Q$ 
are real numbers,
we always have a freedom to replace them by
\begin{equation}
x_P\rightarrow e^{i2\pi n_P}x_P,\hspace{1em}
x_Q\rightarrow e^{i2\pi n_Q}x_Q,
\end{equation}
with integers $n_P$ and $n_Q$.
Since we have no phase transition on the ``path'' of the extended 
model space
$x_P=e^{i2\pi t}|x_P|$ ($0\leq t\leq 1$),
this replacement should not shift the value of
$\theta$ up to modulus $2\pi$.
Thus, $c$ must be an integer.
Consequently, the expression (\ref{c-toy}) requires $\rho$
to be an integer.
When $\rho$ is an odd integer, $g$ must be an even integer
and {\it vice versa}.

These results seem to be inconsistent, 
because we are allowed to have arbitrary number $(g)$ of the chiral
quarks, and also allowed to assign the weights $\alpha$ and $\beta$
to arbitrary positive values. 
Furthermore, this evaluation (\ref{det[naive]})
gives the $\varphi_{\rm PQ}$ dependence
to $\theta$.
From (\ref{theta-toy-tree}), we find
\begin{equation}
\theta=-g\varphi_{\rm PQ}+[\varphi_{\rm PQ}\mbox{-independent terms}].
\end{equation}
This equation states that, once the $SU(1,1)$ symmetry 
is spontaneously broken 
and the chiral quarks are generated $(g\neq 0)$,
the $U(1)_{\rm PQ}$ symmetry becomes an anomalous symmetry
although it is originally a vector symmetry.
If we accept this result, the N-G boson $G^0$ becomes an axion.
\cite{ref:axion}

\subsection*{\it Alternative evaluation}

It is reasonable to suspect that the expression (\ref{det[naive]})
is incorrect because ${\cal M}_{\rm tree}$ is an
infinite-dimensional matrix. 
The (1,2)-entry, (2,1)-entry, and (2,2)-entry sub-matrices of 
${\cal M}_{\rm tree}$ will contribute to $\theta$ by some amount and
modify (\ref{theta-toy-tree}) so that the $U(1)_{\rm PQ}$ symmetry is
exactly maintained. The potentially complex quantities contained in
these sub-matrices are $\langle h\rangle$ and $\epsilon$ in $U_m$. From
the expression (\ref{VEVs-phi}), we find that the only candidate that
eliminates the $\varphi_{\rm PQ}$ dependence of $\theta$ in
(\ref{theta-toy-tree}) is the addition of the term $-g 
\arg[\langle h\rangle]$. Because $\langle h\rangle$ always appears in
${\cal M}_{\rm tree}$ by the combination $yU_0\langle h\rangle$, what we
should add must be $-g \arg[yU_0\langle h\rangle]$. The phase of
$\epsilon$ will also modify $\theta$ by some amount. Since we have no
way to determine the factor, we add $gc' \arg[\epsilon]$ with some
unknown real number $c'$ that will depend on $g$.  In principle, we may
not rule out a possibility that some function of the product of $\langle
h\rangle$, 
$\epsilon$, and $\langle\psi^F_{-g}\rangle$ that has zero total weight
contributes to $\theta$.  Since $\langle h\rangle$ moves under the
$U(1)_{\rm PQ}$ transformation, it cannot join in this product, and the
candidate is limited to the function of
$\langle\psi^F_{-g}\rangle/\epsilon^g$. We omit this possibility because
it seems to be implausible that $\det[{\cal M}_{\rm tree}]$ contains
such a term since ${\cal M}_{\rm tree}$ is decomposable to the product
of the three matrices ${\cal M}_{\rm tree}=A(\epsilon)B(\langle
h\rangle,\langle\psi^F_{-g}\rangle)A(\epsilon)$. The omission of this
term amounts to the assumption that the formula
$\det[XY]=\det[X]\det[Y]$ holds for infinite-dimensional matrices. We
note that even if we incorporate this term, we will see in \S 5
that the phase of $\langle\psi^F_{-g}\rangle/\epsilon^g$ must be
assigned to 0 (mod.$\pi/4$) and it does not give a serious problem to a
discussion of $\theta$. Therefore, under the above assumption, there is
no other way than the addition of the first two terms to modify $\theta$
to recover the $U(1)_{\rm PQ}$ symmetry. 
This leads to
\begin{eqnarray}
&&\theta=\pi-g\left(g-1+c'+2c\right)\varphi_H
+g(g-1+c')\arg[V'/V]\nonumber\\
&&\hspace{2em}+c\left(2\arg[V^F]+\arg[x_Px_Q]\right)
+\arg\det[C_{m,n}^Fb_{m+n}(\rho)],
\end{eqnarray}
from which we find
\begin{equation}
c=-\frac{g-1+c'}{2}.
\end{equation}
So, we have
\begin{equation}
\theta=\pi+c\left(-2g\arg[V'/V]+2\arg[V^F]+\arg[x_Px_Q]
\right)+\arg\det[C_{m,n}^Fb_{m+n}(\rho)].
\label{theta-toy-tree-1}
\end{equation}
Notice that the factor $yU_0\langle h\rangle$ in the (1,1)-entry
sub-matrix of (\ref{toy-mass-matrix}) is completely 
 eliminated in this expression
by the possible contributions from the (1,2), (2,1), and (2,2)-entry
sub-matrices, which are related to the superheavy quarks.
Also, $V^F$, which represents the masses of the superheavy quarks,
appears in (\ref{theta-toy-tree-1}) to cancel the weight of $V'/V$.
These results show that the superheavy particles play a significant
role in $\theta$.
They never decouple from $\theta$.

Although we do not have reliable methods of determining the value of $c$,
it must be an integer depending on $g$.
In this alternative evaluation, the N-G boson $G^0$ is not an axion.
It should be an exactly massless pseudo-scalar particle.
The result (\ref{theta-toy-tree-1}) does not impose any constraint
on the number of the chiral quarks nor on the values of the weights.

\subsection*{{\it Coupling of} $\langle\psi_0^{\prime F}\rangle$}

Let us finally add the coupling 
$(x'_PP_\alpha\bar P_{-\alpha}+x'_Q\bar Q_\beta Q_{-\beta})\Psi^{\prime F}$
with the VEV $\langle\psi_0^{\prime F}\rangle$
to the superpotential (\ref{W-toy}).
Then, the tree level mass terms (\ref{toy-mass-term}) are modified
in the form $\bar{\cal Q}{\cal M}{\cal Q}$ with
\begin{eqnarray}
&&\bar{\cal Q}=\{\bar q_{\beta+i}|\bar p_{-\alpha-j}\},\hspace{1em}
{\cal Q}=\{p_{\alpha+k}|q_{-\beta-l}\},
\hspace{1.5em}i,j,k,l=0,1,2,3,\cdots,\\
&&{\cal M}=\left(\begin{array}{c|c}y\hspace{.1em}C_{i,k}^FU_{i+k}
\langle h\rangle & 
 x_QD_l^Q\langle\psi^F_{-g}\rangle \delta_{i,l+g}
+x'_QD_l^{\prime Q}\langle\psi_0^{\prime F}\rangle\delta_{i,l}\\
\hline
x_PD^P_j\langle\psi^F_{-g}\rangle\delta_{j+g,k}
+x'_PD_j^{\prime P}\langle\psi_0^{\prime F}\rangle\delta_{j,k}
  & 0
\end{array}
\right).\nonumber\\
\label{toy-mass-matrix-1}
\end{eqnarray}
We deform ${\cal M}$ to ${\cal M}'$ by eliminating
$\langle\psi^{\prime F}_0\rangle$
by using $\langle\psi^F_{-g}\rangle$ 
so that ${\cal M}$ and ${\cal M}'$ have the same determinant.
The result is 
\begin{eqnarray}
&&{\cal M}'
=\left(\begin{array}{c|c}y\hspace{.1em}
\sum_{r,s=0}^\infty (-\epsilon_q)^r(-\epsilon_p)^s
b^q_{ir}(\beta)b^p_{ks}(\alpha)
C_{i+gr,k+gs}^FU_{i+gr+k+gs}
\langle h\rangle & 
 x_QD_l^Q\langle\psi^F_{-g}\rangle \delta_{i,l+g}
\\
\hline
x_PD^P_j\langle\psi^F_{-g}\rangle\delta_{j+g,k}
  & 0
\end{array}
\right),\nonumber\\
\label{toy-mass-matrix-2}
\end{eqnarray}
where
\begin{equation}
\epsilon_p
=\frac{x'_P}{x_P}
\frac{\langle\psi_0^{\prime F}\rangle}
{\langle\psi^F_{-g}\rangle},\hspace{1em}
\epsilon_q
=\frac{x'_Q}{x_Q}
\frac{\langle\psi_0^{\prime F}\rangle}
{\langle\psi^F_{-g}\rangle},\hspace{1em}
b^p_{ks}(\alpha)=\prod_{m=0}^{s-1}\frac{D^{\prime P}_{k+gm}}{D^P_{k+gm}},
\hspace{1em}
b^q_{ir}(\beta)=\prod_{m=0}^{r-1}\frac{D^{\prime Q}_{i+gm}}{D^Q_{i+gm}},
\end{equation}
This matrix has almost the same structure as (\ref{toy-mass-matrix}).
When we perform the naive evaluation of $\theta$,
the difference is only the replacement of the mass matrix of the $g$
generations of chiral quarks, except for the absence of the normalization
factors $U^q_m(<1)$ and $U^p_n(<1)$ of the chiral quarks.
Their absence is understandable because
the elimination
of $\langle\psi_0^{\prime F}\rangle$ decreases the masses of the
superheavy quarks but $\det[{\cal M}]$ 
must be unchanged.

Let us evaluate $\theta$ based on the expression (\ref{toy-mass-matrix-2}).
The naive evaluation gives
\begin{eqnarray}
&&\theta_{\rm naive}=\pi-g(g-1+\rho+2c)\varphi_H-g\varphi_{\rm PQ}
+g(g-1)\arg[V'/V]+g\arg[v]\nonumber\\
&&\hspace{3.0em} +c(2\arg[V^F]+\arg[x_Px_Q])\nonumber\\
&&\hspace{3.0em} +\arg\det\left[y\sum_{r,s=0}^{\infty}
(-\epsilon_q\epsilon^g)^r(-\epsilon_p\epsilon^g)^s
b^q_{mr}(\beta)b^p_{ns}(\alpha)
C^F_{m+gr,n+gs}U_0b_{m+n+g(r+s)}(\rho)
\right].\nonumber\\
\label{naive-2}
\end{eqnarray}
To obtain the alternative evaluation of $\theta$, 
we must examine the detailed
structure of the (1,1)-entry matrix of (\ref{toy-mass-matrix-2}):
\begin{eqnarray}
&&{\cal M}_{(1,1)}^{ik}=y\hspace{.1em}
\sum_{r,s=0}^\infty (-\epsilon_q)^r(-\epsilon_p)^s
b^q_{ir}(\beta)b^p_{ks}(\alpha)
C_{i+gr,k+gs}^FU_{i+gr+k+gs}
\langle h\rangle\nonumber\\
&&\hspace{3.3em}=y\hspace{0.1em}U_0\langle h\rangle\epsilon^{i+k}
\sum_{r,s=0}^\infty (-\epsilon_q\epsilon^g)^r(-\epsilon_p\epsilon^g)^s
b^q_{ir}(\beta)b^p_{ks}(\alpha)
C_{i+gr,k+gs}^Fb_{i+k+g(r+s)}(\rho).\nonumber\\
\label{M(1,1)}
\end{eqnarray}
The elimination of the $\varphi_{\rm PQ}$ dependence from
(\ref{naive-2})
is accomplished only by adding a term 
$-g\arg[yU_0\langle h\rangle]$ as we did.
Also, because $\epsilon^{i+k}$ is an overall factor,
its possible effect is the addition of the term
$gc'\arg[\epsilon]$.
The remaining matrix in (\ref{M(1,1)}) contains 
potentially complex extra quantities
$\epsilon_q\epsilon^g$ and $\epsilon_p\epsilon^g$. 
Since this matrix is symmetric under the interchange
\begin{equation}
(i,\beta,\epsilon_q\epsilon^g)\leftrightarrow
(k,\alpha,\epsilon_p\epsilon^g)
\end{equation}
owing to the symmetry relation (\ref{C-sym}) of the C-G coefficient
$C^F_{i,k}$, 
its possible effect to $\theta$ should be through the term
$\arg[f^{\beta,\alpha}(\epsilon_q\epsilon^g,\epsilon_p\epsilon^g)]$,
where $f^{\beta,\alpha}(x,y)=f^{\alpha,\beta}(y,x)$ 
is some real function of $x$ and $y$.
Thus, $\theta$ should take the form
\begin{eqnarray}
&&\theta=\pi+c\left(-2g\arg[\epsilon]+2\arg[\langle\psi_{-g}^F\rangle]
+\arg[x_Px_Q]
\right)+\arg[f^{\beta,\alpha}(\epsilon_q\epsilon^g,\epsilon_p\epsilon^g)]
\nonumber\\
&&\hspace{2em}+\arg\det\left[\sum_{r,s=0}^{\infty} 
(\epsilon_q\epsilon^g)^r(\epsilon_p\epsilon^g)^s
b^q_{mr}b^p_{ns}C^F_{m+gr,n+gs}b_{m+n+g(r+s)}(\rho)
\right].
\label{theta-toy-tree-2}
\end{eqnarray}
One may doubt about the prescription adopted to eliminate
$\langle\psi_0^{\prime F}\rangle$.
The unterminated series of $\langle\psi_0^{\prime F}\rangle$ may 
still contribute to $\theta$ with a term proportional 
to $\arg[\langle\psi_0^{\prime F}\rangle]$.
However, the coefficient $c_0$ of 
$\arg[\langle\psi_0^{\prime F}\rangle]$ must be an integer,
and the limit $x'_P,x'_Q\rightarrow 0$ must reproduce $c_0\rightarrow 0$.
Since we have no phase transition in this limit,
we should have $c_0=0$.
Notice that (\ref{theta-toy-tree-2}) precisely reproduces
(\ref{theta-toy-tree-1}) in this limit.

\section{Tree level analysis of the strong CP problem}

Now that we have obtained sufficient knowledge
on what happens when the $SU(1,1)$ symmetry is spontaneously
broken, let us return to the main subject.
The total superpotential of the model
\begin{equation}
W=W_{\rm quark}+W_{\rm higgs}+W_{\rm M}+W_{\rm Y}+W[\mbox{finite dim.}]
\label{total-W}
\end{equation}
respects the $U(1)_{\rm PQ}$ symmetry.
The supersymmetry breaking terms also respect this symmetry.
This symmetry is spontaneously broken at the energy scale 
$\sqrt{m_{\rm SUSY}M}$ by the VEVs of $r$ and $\bar r$.
The VEVs of the higgses $h$ and $h'$ also break this symmetry
at the scale $m_{\rm SUSY}$.
If we accept the naive evaluation (\ref{theta-toy-tree}),
the $U(1)_{\rm PQ}$
symmetry is now an anomalous symmetry because we have three generations
of the chiral quarks.
This means that the associated N-G boson $G^0$ is 
an invisible axion.\cite{ref:invisible-axion}
Therefore, $G^0$ will solve the strong CP problem
in the present model.
We do not, however, accept this solution,
since it seems to lead to some inconsistency 
observed in the previous section.
We make a trial of solving the problem based on the alternative
evaluation of $\theta$ presented in (\ref{theta-toy-tree-1})
and (\ref{theta-toy-tree-2}).

Let us evaluate the value of $\theta$ in the model
coming from the diagonalization of the quark mass matrices.
Remember that the original value of $\theta$ is $\theta_0=0$.
The relevant part of the superpotential for the subject 
is the mass operators of the quarks:
\begin{equation}
W_{\rm mass}=W_{\rm quark}+W_{\rm Y}.
\end{equation}
We introduce a notation
\begin{equation}
Q_\alpha=\left(\begin{array}{c}U_\alpha^q\\ D_\alpha^q\end{array}
\right),\hspace{1.5em}
U_\alpha^q=\{u_\alpha^q,u_{\alpha+1}^q,\cdots\},\hspace{1em}
D_\alpha^q=\{d_\alpha^q,d_{\alpha+1}^q,\cdots\}.
\end{equation}
We examine the Type-II scheme (\ref{Yukawa-II}) for $W_{\rm Y}$,
which is simpler than the Type-I scheme.
The up-type and down-type quark mass operators are
expressed as
\begin{eqnarray}
&&W^u_{\rm mass}=(x_QU^q_\alpha\bar U^q_{-\alpha}
+x_U\bar U_\beta U_{-\beta})\Psi^F
+(x'_QU^q_\alpha\bar U^q_{-\alpha}
+x'_U\bar U_\beta U_{-\beta})\Psi'^F\nonumber\\
&&\hspace{4em}+y_U(\bar U_\beta U^q_\alpha H^2_{-\rho}
+U_{-\beta}\bar U^q_{-\alpha}\bar H^2_\rho),
\label{up-mass-operator}\\
&&W^d_{\rm mass}=(x_QD^q_\alpha\bar D^q_{-\alpha}
+x_D\bar D_\gamma D_{-\gamma})\Psi^F+
(x'_QD^q_\alpha\bar D^q_{-\alpha}
+x'_D\bar D_\gamma D_{-\gamma})\Psi'^F\nonumber\\
&&\hspace{4em}-y_D(\bar D_\gamma D^q_\alpha H'^1_{-\sigma}
+D_{-\gamma}\bar D^q_{-\alpha}\bar H'^1_\sigma),
\label{down-mass-operator}
\end{eqnarray}
with the VEVs
\begin{equation}
\langle\Psi^F\rangle=\langle\psi^F_{-3}\rangle,~
\langle\Psi'^F\rangle=\langle\psi'^F_{0}\rangle,~
\langle H_{-\rho}^2\rangle=\langle h^2_{-\rho-i}\rangle
=\langle h^2\rangle U_i,~
\langle H^{\prime 1}_{-\sigma}\rangle
=\langle h^{\prime 1}_{-\sigma-i}\rangle
=\langle h^{\prime 1}\rangle U'_i.
\end{equation}
When we set $x_Q=x_U=x_D=0$, all quarks become superheavy,
but when we set $x'_Q=x'_U=x'_D=0$, we have three generations of the quarks.
This means that there is a critical value $\epsilon_{\rm cr}$ 
for the ratios
\begin{equation}
\epsilon_q=\frac{x_Q^\prime}{x_Q}
\frac{\langle\psi_0^{\prime F}\rangle}
{\langle\psi_{-3}^F\rangle},\hspace{1.5em}
\epsilon_u=\frac{x_U^\prime}{x_U}
\frac{\langle\psi_0^{\prime F}\rangle}
{\langle\psi_{-3}^F\rangle},\hspace{1.5em}
\epsilon_d=\frac{x_D^\prime}{x_D}
\frac{\langle\psi_0^{\prime F}\rangle}
{\langle\psi_{-3}^F\rangle}.
\label{epsilon-q-u-d}
\end{equation}
The critical ratio $\epsilon_{\rm cr}$ is derived from the normalizable
condition $\sum_{i=0}^\infty|U_{m\ i}^f|^2<\infty$ $(f=q,u,d)$
for the mixing coefficients (\ref{q-mixing}) of the chiral modes,
which requires ${\rm lim}_{i\rightarrow\infty}|U_{m\ i+3}^f|/|U_{m\ i}^f|<1$.
This gives \cite{ref:su11-2}
\begin{equation}
\epsilon_{\rm cr}=\sqrt{\frac{S^F(S^F-1)(S^F-2)}{(S^F+3)(S^F+2)(S^F+1)}}\ .
\end{equation}
To generate the three generations of $q_m$, $\bar u_m$, and $\bar d_m$,
all ratios in (\ref{epsilon-q-u-d}) must satisfy
\begin{equation}
|\epsilon_q|, |\epsilon_u|, |\epsilon_d| < \epsilon_{\rm cr}.
\end{equation}

If we set $x'_Q=x'_U=x'_D=0$, the mass operators (\ref{up-mass-operator}) 
and (\ref{down-mass-operator}) reduce to the duplication of the 
toy model (\ref{W-toy}).
Since the coefficient $c$ in (\ref{theta-toy-tree-1}) is an integer
and it should be 
independent of the weights $\alpha$, $\beta$,
and $\gamma$, it is sure that the up-type quarks and the down-type
quarks take the common value.
Thus, the diagonalization of the quark mass matrices will generate 
$\theta_{\rm tree}$ of the amount
\begin{eqnarray}
&&\theta^0_{\rm tree}=
c\left(-6\arg[\epsilon]+2\arg[\langle\psi^F_{-3}\rangle]+\arg[x_Qx_U]
-6\arg[\epsilon']+2\arg[\langle\psi^F_{-3}\rangle+\arg[x_Qx_D]]
\right)
\nonumber\\
&&\hspace*{2.7em}
+\arg\det[C_{m,n}^{\beta,\alpha}b_{m+n}(\rho)]
+\arg\det[C_{m,n}^{\gamma,\alpha}b'_{m+n}(\sigma)]
\label{theta-tree}
\end{eqnarray}
up to modulus $2\pi$.

When we turn on the couplings $x'_Q$, $x'_U$, and $x'_D$,
all chiral quarks are realized as the superpositions of the infinite
number of the components of each $SU(1,1)$ multiplet.
Then, $\theta^0_{\rm tree}$ is modified by
$x'_Q\langle\psi_0^{\prime F}\rangle$, 
$x'_U\langle\psi_0^{\prime F}\rangle$, and
$x'_D\langle\psi_0^{\prime F}\rangle$.  
From the final result (\ref{theta-toy-tree-2}) of the toy model,
we obtain $\theta_{\rm tree}$, for the Type-II scheme, 
in the form
\begin{eqnarray}
&&\theta_{\rm tree}=
c\left(-\arg[\epsilon_q\epsilon^3]-\arg[\epsilon_q\epsilon^{\prime 3}]
-\arg[\epsilon_u\epsilon^3]-\arg[\epsilon_d\epsilon^{\prime 3}]
+4\arg[\langle\psi_0^{\prime F}\rangle]+\arg[x'_Ux'_D]
\right)\nonumber\\
&&\hspace{2em}
+\arg[f^{\beta,\alpha}
(\epsilon_u\epsilon^3,\epsilon_q\epsilon^3)]
+\arg[f^{\gamma,\alpha}
(\epsilon_d\epsilon^{\prime 3},\epsilon_q\epsilon^{\prime 3})]
+\arg\det[A_u^{mn}]
+\arg\det[A_d^{mn}].\nonumber\\
\label{theta-tree-II}
\end{eqnarray} 

Let us next examine the Type-I scheme (\ref{Yukawa-I}) for $W_{\rm Y}$.
When we extend the toy model (\ref{W-toy})
so that it is applicable in this scheme,
the matrix (\ref{M(1,1)}) is replaced by 
\begin{eqnarray}
&&
y\hspace{0.1em}U_0\langle h\rangle\epsilon^{i+k}
\left(
\sum_{r,s=0}^\infty (-\epsilon_q\epsilon^g)^r(-\epsilon_p\epsilon^g)^s
b^q_{ir}(\beta)b^p_{ks}(\alpha)
C_{i+gr,k+gs}^F(0)b_{i+k+g(r+s)}(\rho)\right.\nonumber\\
&&\hspace{2.5em}
-r_F\sum_{r,s=0}^\infty\theta_{i+k+g(r+s),\Delta} 
(-\epsilon_q\epsilon^g)^r(-\epsilon_p\epsilon^g)^s\nonumber\\
&&\hspace{3.5em}
\times~b^q_{ir}(\beta)b^p_{ks}(\alpha)
C_{i+gr,k+gs}^F(\Delta)b_{i+k+g(r+s)-\Delta}(\rho)R_{i+gr+k+gs-\Delta}(\rho)
\bigg).
\label{M(1,1)-I}
\end{eqnarray}
Notice that the matrix in the parenthesis is symmetric under the
interchange
\begin{equation}
(i,\beta,\epsilon_q\epsilon^g,r_F)\leftrightarrow
(k,\alpha,\epsilon_p\epsilon^g,(-1)^\Delta r_F)
\end{equation}
Therefore, following the same consideration given at the end of the
previous section,
we obtain the expression of $\theta_{\rm tree}$ for the Type-I
scheme in the form
\begin{eqnarray}
&&\theta_{\rm tree}=
c\left(-\arg[\epsilon_q\epsilon^3]-\arg[\epsilon_q\epsilon^{\prime 3}]
-\arg[\epsilon_u\epsilon^3]-\arg[\epsilon_d\epsilon^{\prime 3}]
+4\arg[\langle\psi_0^{\prime F}\rangle]+\arg[x'_Ux'_D]
\right)\nonumber\\
&&\hspace{2em}
+\arg[f^{\beta,\alpha}
(\epsilon_u\epsilon^3,\epsilon_q\epsilon^3,r_U)]
+\arg[f^{\gamma,\alpha}
(\epsilon_d\epsilon^{\prime 3},\epsilon_q\epsilon^{\prime 3},r_D)]
\nonumber\\
&&\hspace{2em}
+\arg\det[Y_u^{mn}]
+\arg\det[Y_d^{mn}],
\label{theta-tree-full-00}
\end{eqnarray} 
where $f^{\beta,\alpha}(x,y,z)$ is a real function of $x$, $y$, and $z$
satisfying the relation
\begin{equation}
f^{\beta,\alpha}(x,y,z)=f^{\alpha,\beta}(y,x,(-1)^\Delta z).
\label{f-symmetry-I}
\end{equation}
The expression (\ref{theta-tree-full-00}) covers both of the Type-I
and Type-II $(r_U=r_D=0)$ schemes.
Remember that $c$ is some real integer, although we cannot determine its
value.

Now, we proceed to the main subject of the problem.
We must search for a solution to suppress $\theta_{\rm tree}$
in (\ref{theta-tree-full-00}) in a natural manner.
What we should attain is $\theta_{\rm tree}=0$ (mod. $\pi$).
We note that the value $\theta=\pi$ does not bring the strong CP problem.
Let us first investigate (\ref{theta-tree-II})
in detail for the Type-II scheme.
From their expressions (\ref{A-u}) and (\ref{A-d}),
we find remarkable facts.
Although the matrices $A_u^{mn}$ and $A_d^{mn}$ consist of the
infinite number of terms, 
each term contains a potentially complex quantity only by the 
factor $(-\epsilon_u\epsilon^3)^r(-\epsilon_q\epsilon^3)^s$
or $(-\epsilon_d\epsilon^{\prime 3})^r(-\epsilon_q\epsilon^{\prime 3})^s$.
If these quantities were complex numbers,
it will be hard to expect a miracle occurs so that 
$\det[A_u^{mn}]$ and $\det[A_d^{mn}]$ have the natural phases
of realizing $\theta_{\rm tree}=0$.
That is, the phase assignments
\begin{equation}
\arg[\epsilon_q\epsilon^3]=0,\ \arg[\epsilon_u\epsilon^3]=0,\ 
\arg[\epsilon_q\epsilon^{\prime 3}]=0,\ 
\arg[\epsilon_d\epsilon^{\prime 3}]=0\ \ 
(\mbox{mod}.\ \pi)
\label{arg-1}
\end{equation}
are indispensable.
Then, $\det[A_u^{mn}]$ and $\det[A_d^{mn}]$ are real.
What is surprising is that these phase assignments 
ensure almost all terms except $\arg[\langle\psi_0^{\prime F}\rangle]$
in (\ref{theta-tree-II}) to vanish up to modulus $\pi$.
Since $\epsilon_f$'s $(f=q,u,d)$ have the common phase up to $\pi$,
the net conditions are
\begin{equation}
\arg[\epsilon_q\epsilon^3]=0,\ 
\arg[\epsilon_q\epsilon^{\prime 3}]=0 \ 
(\mbox{mod}.\ \pi).
\label{arg-1-net}
\end{equation}
The remaining subject is the phase of 
$\langle\psi_0^{\prime F}\rangle$.
The sufficient condition of
eliminating this contribution is
\begin{equation}
\arg[\langle\psi_0^{\prime F}\rangle]=0 \ (\mbox{mod}.\ \pi/4).
\label{arg-2}
\end{equation}
Thus, the conditions (\ref{arg-1-net}) and (\ref{arg-2})
realize
\begin{equation}
\theta_{\rm tree}=0\ \ (\mbox{mod}.\ \pi).
\label{theta-II}
\end{equation}
Notice that the conditions (\ref{arg-1-net})
fix the relative phase of $\epsilon$ and $\epsilon'$
in the way
\begin{equation}
\arg[\epsilon/\epsilon^{\prime}]=0\ \ (\mbox{mod}.\ \pi/3).
\label{epsilon/epsilon'}
\end{equation}
It will be evident that the integer 3 in the denominator
of $\pi/3$ in (\ref{epsilon/epsilon'})
is a consequence of the fact that we have 
three generations of the chiral quarks.

Next, we investigate the Type-I scheme.
It will be obvious that the conditions
(\ref{arg-1-net}) and (\ref{arg-2}) are also indispensable for realizing
$\theta_{\rm tree}=0$ in a natural way.
However, the situation is somewhat complicated because this scheme
contains the additional quantities $r_U$ and $r_D$.
The expression of $\theta_{\rm tree}$ is now
\begin{eqnarray}
&&\theta_{\rm tree}=
\arg\det [Y_u^{mn}]+\arg\det [Y_d^{mn}]\nonumber\\
&&\hspace{2em}+\arg[f^{\beta,\alpha}
(\epsilon_u\epsilon^3,\epsilon_q\epsilon^3,r_U)]
+\arg[f^{\gamma,\alpha}
(\epsilon_d\epsilon^{\prime 3},\epsilon_q\epsilon^{\prime 3},r_D)]
\ \ (\mbox{mod}.\ \pi),
\label{theta-I}
\end{eqnarray}
with real $\epsilon_f\epsilon^3$
and $\epsilon_f\epsilon^{\prime 3}$.
Since we cannot expect a delicate cancellation between two nontrivial 
phases of the first two terms in (\ref{theta-I}), 
we are forced to require
\begin{equation}
\arg\det [Y_u^{mn}]=0,\ \ \ 
\arg\det [Y_d^{mn}]=0\ \ \ (\mbox{mod}.\ \pi).
\end{equation}
The obvious candidate for meeting these requirements is
\begin{equation}
\mbox{Option-1}:\hspace{2em}
\arg[r_U]=0,\ \arg[r_D]=0\ \ (\mbox{mod}.\ \pi).
\label{Option-1}
\end{equation}
Then, the third and the fourth terms in (\ref{theta-I}) 
also vanish, and we have $\theta_{\rm tree}=0$ (mod. $\pi$).

Another candidate, which is rather subtle, is
\begin{equation}
\ \ \mbox{Option-2}:\hspace{2em} 
\arg[r_U]=\frac{\pi}{2},\ 
\arg[r_D]=\frac{\pi}{2}\ \ (\mbox{mod}.\ \pi).
\label{Option-2}
\end{equation}
In this case, we need the additional conditions:
\begin{equation}
\alpha=\beta=\gamma,\hspace{2em}
\frac{x^\prime_Q}{x_Q}=\frac{x^\prime_U}{x_U}=\frac{x^\prime_D}{x_D},
\hspace{2em}\Delta,\ \Delta^\prime=[\mbox{odd integer}].
\label{universal}
\end{equation}
The first two conditions give
\begin{equation}
b^q_{mr}(\alpha)=b^u_{mr}(\beta)=b^d_{mr}(\gamma),\hspace{2em}
\epsilon_q=\epsilon_u=\epsilon_d .
\end{equation}
Then, the third condition implies
that the matrices $A_u^{mn}$ and $A_d^{mn}$
are symmetric but $B_u^{mn}$ and $B_d^{mn}$ are anti-symmetric
owing to the symmetry relation (\ref{C-sym}).
Consequently, the phase assignment (\ref{Option-2}) realizes the 
hermitian matrices for $Y_u^{mn}$ and $Y_d^{mn}$ whose determinants
are real.
What is unexpected is that the conditions (\ref{Option-2})
and (\ref{universal}) also render 
the third and the fourth terms in (\ref{theta-I})
to vanish separately.
This is because the symmetry relation (\ref{f-symmetry-I})
implies, under the odd value of $\Delta$,
that the function 
$f^{\alpha,\alpha}(\epsilon_q\epsilon^3,\epsilon_q\epsilon^3,r_U)$
is pure real when $r_U$ is pure imaginary: 
\begin{equation}
f^{\alpha,\alpha}
(\epsilon_q\epsilon^3,\epsilon_q\epsilon^3,r_U)=
f^{\alpha,\alpha}
(\epsilon_q\epsilon^3,\epsilon_q\epsilon^3,-r_U)=
f^{\alpha,\alpha}
(\epsilon_q\epsilon^3,\epsilon_q\epsilon^3,r_U)^*.
\end{equation}
Of course, we also have the mixed candidates. That is,
the up-type quarks
take the Option-1 phase assignment and the down-type quarks
take the Option-2 assignment and {\it vice versa}.

The universal quark assignment (\ref{universal}) for the Option-2 case
seems to invoke the grand unified gauge group
$G\supset SU_3\times SU_2\times U_1$.
\cite{ref:SU(5)GUT,ref:SO(10)GUT,ref:E6GUT,ref:GUT-rev}
When $G=SU(5)$, it is not possible to satisfy all requirements in 
(\ref{universal}), and the case is limited to the mixed candidate.
One may imagine that $G=SO(10)$
easily fills (\ref{universal})
by simply assigning $\Psi^F$ and $\Psi^{\prime F}$ to 
the common representation of $SO(10)$.
This assignment, however, generates the massless right-handed neutrinos.
As far as we have examined, it seems to be hard to eliminate
the right-handed neutrinos as an illusion \cite{ref:su11-3}
without disturbing (\ref{universal}).
The possibility is open for $G\supset E_6$.

It is worth noting that, when the up-type and the down-type quarks
take the Type-II scheme and/or the Type-I scheme with the Option-1
phase assignment,
the phases residing in the Yukawa coupling
matrices (\ref{y-u}) and (\ref{y-d}) are only in $\epsilon$
and $\epsilon'$.
This means that 
the origin of the phases in the CKM matrix is
solely the relative phase of $\epsilon$ and $\epsilon'$,
which has been fixed by (\ref{epsilon/epsilon'}) to be 
$\delta\equiv\arg[\epsilon/{\epsilon'}]=n\pi/3$ with some integer $n$.
Then, the CKM matrix is restricted to the form
\begin{equation}
V_{\rm CKM}=O_u^TPO_d ,\hspace{2em}
P\equiv\left(\begin{array}{ccc}e^{2i\delta}&0&0\\0&e^{i\delta}&0\\
0&0&1\end{array}\right),\hspace{2em}
\delta=n\pi/3,
\label{CKM-II,I-1}
\end{equation}
where $O_u$ and $O_d$ are the real orthogonal matrices, 
standing on the right when 
the ``reversed'' $(m,n=2,1,0)$ coupling matrices
$y_u^{mn}$ and $y_d^{mn}$ with real and positive $\epsilon$ and $\epsilon'$
are diagonalized, respectively.

\section{Consideration on the radiative corrections}

In the toy model analysis given in the \S 4, we found 
that $\theta$ does not receive the radiative
corrections of the low-energy physics.
In the MSSM, 
we have two higgses $h$ and $h'$, each of which has
its own Yukawa coupling matrix.
Through the supersymmetry breaking terms,
$h^*$ couples to the down-type quarks and $h'^*$ to the up-type quarks.
At the one loop level, this is achieved by the squark-higgsino
loop.
As a result, the phase structures of the mass matrices 
${\cal M}_u^{mn}$ and ${\cal M}_d^{mn}$ are disturbed.
This seems to indicate that $\theta$ will receive the 
uncontrollable radiative
correction $\Delta\theta$.

We must, however, remember that the superheavy quarks do not decouple
from $\theta$. 
We found that the factor $yU_0\langle h\rangle$ in the tree level
mass matrix of the chiral quarks 
was completely eliminated in the final expression of $\theta$
presented in (\ref{theta-toy-tree-2})
by the contributions from masses of the superheavy quarks.
In the present model, 
all entries of the mass matrix receive the nontrivial
vertex corrections.
This means that we must take into account all radiative corrections
not only from the chiral particles but also from the superheavy particles.
It is not unreasonable to expect that both effects 
of the radiative corrections exactly cancel each other
when $\theta=0$ is naturally realized at the tree level.

Let us first consider the factors $y_U\langle h \rangle$ appearing 
in the mass matrix ${\cal M}^{ij}_{\rm tree}$ of the up-type
quarks. Each of them receives a very complicated vertex correction and
is modified to the order $m_{\rm SUSY}$ quantity
$X^{ij}=y_U\langle h \rangle + c^{ij}y_U(y_D)^2\langle h'
\rangle^* + \cdots$. However, their $U(1)_{\rm PQ}$ transformation
property is simply determined, by that of the quarks multiplied to this
mass matrix, to be $X^{ij} \rightarrow e^{-i\varphi_{\rm PQ}} X^{ij}$ (a
coefficient $c^{ij}$ is an appropriate function of the relevant VEVs
that adjust the $U(1)_H$ and $U(1)_{\rm PQ}$ charges). 
The exact $U(1)_{\rm PQ}$ symmetry at the quantum level, therefore,
requires that the contribution of $X^{ij}$ to $\theta$ must be 
canceled between the chiral and the superheavy quarks in the same way as
we found at the tree level. 
The radiatively induced new mass operator 
$U{\widetilde {\cal M}}\bar U^q$ is
suppressed to the negligibly small order 
${\widetilde {\cal M}} \sim m_{\rm SUSY} (m_{\rm SUSY}/M)^2$ 
and thus it does not disturb this cancellation though 
${\widetilde {\cal M}}$ has $Q_{PQ}=1$. 
Because the exact $U(1)_{\rm PQ}$ symmetry owes its realization fully to
the fact that the model is based on the framework of the vectorlike
gauge theory, it will be reasonable to expect that the true origin of
this cancellation is the vectorlike feature of the model rather than the
exact $U(1)_{\rm PQ}$ symmetry.
If this consideration is correct, the vertex corrections to the elements
of ${\cal M}^{ij}_{\rm tree}$ that contain $\langle \psi_{-3}^F\rangle$
and $\langle \psi_{0}^F\rangle$ and ${\widetilde {\cal M}}$ will be also
canceled in $\theta$. 
Therefore, the possible source of the modification to $\theta_{\rm
tree}=0$ is the radiative effects to the relative phases of the VEVs of
$\Psi$'s. If these phases at the tree-level are determined so that they
depend continuously on the coupling constants of the model, there will
be no reason to expect that these relative phases do not receive
radiative corrections, and we will have sizable $\Delta\theta$. 
However, when the VEVs $\langle \Psi\rangle$'s are determined through
their superpotential $W[\mbox{finite-dim.}]$ so that they have natural
phases given in (\ref{natural-phase}) and realize $\theta_{\rm tree}=0$,
it seems to be probable that $\theta_{\rm tree}=0$ will not be affected
by the radiative corrections as we will argue in the following.

As an illustration, let us consider the $\lambda\phi^4$ theory
with negative mass-square.
This theory has two degenerate vacua 
\begin{equation}
\langle\phi\rangle_1=-\langle\phi\rangle_2\equiv v_0.
\label{phi=v_0}
\end{equation}
Although the magnitude of $v_0$ receives the radiative correction,
the relation (\ref{phi=v_0}) is protected by the symmetry
$\phi\rightarrow -\phi$ of the theory.
The directions of the vacua from the origin $(\phi=0)$ are determined
by the tree level analysis.
In general, when a theory has some discrete symmetry $D$,
and $D$ is spontaneously broken,
the degenerate vacua should take definite directions in the 
field-space of the theory specified by the symmetry of the theory.

The present model has the P-C-T-invariance at the fundamental level,
and all of these discrete symmetries are spontaneously broken
by the VEVs of $\Psi$'s.
We have realized $\theta=0$ at the tree level in the natural manner, 
at least for the Type-I scheme with Option-1 phase assignment
and the Type-II scheme, which depends only on the directions (phases)
of the VEVs and does not depend on the detailed values 
of the coupling constants of the model. 
Even in the Type-I scheme with Option-2 phase assignment,
when the second relation in (\ref{universal}) is protected
by the gauge symmetry of the grand unified theory,
the realization $\theta=0$ is natural.
Remember that $\theta=0$ is a minimum point of the vacuum energy.
Suppose $\theta_{\rm tree}=0$ receives the radiative corrections.
Then, there will be the direction of the VEVs that realizes $\theta=0$
in the vicinity of the original direction.
However, this direction must depend on the coupling constants
of the model to cancel the radiative corrections.
It seems to be very unbelievable that the direction of the VEVs,
which spontaneously 
realizes $\theta=0$, depends on the coupling constants of the model.

One may wonder why $\theta_{\rm tree}=0$ that is realized in the models
based on the spontaneous breakdown of the CP-invariance 
\cite{ref:spontaneous-CP} 
receives the radiative corrections.
The essential reason is that the gluon $\theta$-term 
(\ref{theta-term}) breaks
not only the CP-invariance but also the P-invariance.
The ordinary models with the spontaneous CP-violation 
are based on the framework of the chiral theory, which
explicitly breaks the P-invariance.
Therefore, the emergence of $\theta$ is not a genuine product of the
spontaneous symmetry breaking.
This means that, even if $\theta_{\rm tree}=0$ is achieved in these models,
it is not protected against the radiative corrections 
by the symmetry of the model. 

In the present model, all of the P, C, and T symmetries are spontaneously
broken.
Especially, the spontaneous P breaking generates the chiral 
particles and the superheavy particles.
This means that the original symmetry of the model manifests
its characteristics only when all particles, chiral as well as
superheavy, are taken into account.
It is very hard to show explicitly 
how the radiative corrections
due to the chiral and the superheavy particles cancel each other,
because the model inevitably contains the infinite number of particles.
In fact, the argument presented in this section does not verify the
absence of the radiative correction to $\theta_{\rm tree}=0$.
However, it seems to be quite tempting and leads us to believe that the
present model has a capability of giving $\theta=0$ to the full order of
the quantum effects. 
We would like to leave the attempt on the rigorous proof to the future
study.

\section{Explicit {\boldmath $U(1)_{\rm PQ}$} breaking}

In the previous two sections, we have shown
that the strong CP problem will be solvable
within the natural phase assignment of the VEVs of the 
finite-dimensional multiplets $\Psi$'s.
We remind that the model is invested with the exact $U(1)_{\rm PQ}$
symmetry.
This symmetry is spontaneously broken at the energy scale
$E\simeq\sqrt{m_{\rm SUSY}M}$.
Thus, the model inevitably contains the exactly massless N-G boson
$G^0$, which will mediate a long-range force.
One may worry about this long-range force
since $G^0$ is shared not only with $r$ and $\bar r$ but also with
$h$ and $h^\prime$ with a fraction of the order 
$\langle h\rangle/\langle r\rangle\simeq \sqrt{m_{\rm SUSY}/M}$.
However, $G^0$ will not disturb the Newton's law of the gravitation,
because $G^0$ is a pseudo-scalar particle.
It does not give a sizable force between two massive objects.
Furthermore, the couplings of $G^0$ to quarks and leptons are suppressed
by the factor $\sqrt{m_{\rm SUSY}/M}\simeq 10^{-7}$,
which will be sufficient to clear the present experimental limit.
\cite{ref:PDG}

The $U(1)_{\rm PQ}$ symmetry has played an indispensable role
in deriving the consistent prescription for treating the 
infinite-dimensional matrix.
Nevertheless, we must doubt about this symmetry.
This is an accidental symmetry, which happened to emerge in the process
of the model building to realize the MSSM at the low-energy
by retaining the indispensable couplings
of the matter multiplets.
It will be probable that there are some other couplings that 
explicitly break this symmetry even though they are 
allowed by the $SU(1,1)$
symmetry.
Then, $G^0$ will acquire the mass $m_G$.
The magnitude of $m_G$ should depend on how the $U(1)_{\rm PQ}$
symmetry is explicitly broken.
To estimate the values of $m_G$,
we must identify the relevant couplings.

We search for the candidates for the couplings.
If we limit our considerations to the cubic couplings,
the $SU(1,1)$ invariance gives a rigid restriction
on the possible couplings.
The couplings 
$A_\eta B_\lambda \bar C_{-\zeta}$ and 
$\bar A_{-\eta}\bar B_{-\lambda}C_\zeta$
are allowed only when 
$\zeta-\eta-\lambda$ is a nonnegative integer.
Let us search for the candidate following the power of $S_{\rho+\sigma}$.
The couplings $S^3$ and $\bar RS^2$ are trivially forbidden.
The possible couplings that contain $S^2$ or $\bar S^2$ are
\begin{equation}
\bar R'S^2+R'\bar S^2\hspace{3em}\mbox{with}\hspace{1em}
\rho+\sigma\leq 2/3.
\label{PQ-br-1}
\end{equation}
The candidates containing single $S$ or $\bar S$ are 
\begin{eqnarray}
&&R\bar R'S+\bar RR'\bar S\hspace{3em}\mbox{with}\hspace{1em}
\rho+\sigma\leq 1,
\label{PQ-br-2}
\\
&&\bar HKS+H\bar K\bar S\hspace{3em}\mbox{with}\hspace{1em}
\Delta\geq \rho+\sigma,
\label{PQ-br-3}
\\
&&\bar H'K'S+H'\bar K'\bar S\hspace{2em}\mbox{with}\hspace{1em}
\Delta'\geq \rho+\sigma.
\label{PQ-br-4}
\end{eqnarray}
The couplings that do not contain $S$ nor $\bar S$ are
\begin{eqnarray}
&&\bar HKR+H\bar K\bar R\hspace{3em}\mbox{with}\hspace{1em}
\Delta\geq(\rho+\sigma)/2,
\label{PQ-br-5}
\\
&&\bar H'K'R+H'\bar K'\bar R\hspace{2em}\mbox{with}\hspace{1em}
\Delta'\geq(\rho+\sigma)/2,
\label{PQ-br-6}
\\
&&HH'R'+\bar H\bar H'\bar R'\hspace{2em}\mbox{with}\hspace{1em}
\rho+\sigma=2,
\label{PQ-br-7}
\\
&&RR\bar R'+\bar R\bar RR'\hspace{3.2em}\mbox{with}\hspace{1em}
\rho+\sigma=2.
\label{PQ-br-8}
\end{eqnarray}

It will be obvious that, if the supersymmetry is exact,
$G^0$ keeps its vanishing mass $m_G=0$,
because we have no mass scale other than $M\simeq 10^{16}$GeV.
This means that the sources that give the nonvanishing
mass to $G^0$ are classified into two types.
One is the supersymmetry breaking mass scale
$m_0(\simeq m_{\rm SUSY})$ itself in the $A$-terms.
Another one is through the VEVs induced by the supersymmetry breaking.
The VEVs relevant to the latter are enumerated as
\begin{eqnarray}
&&\langle H^2\rangle=\langle h^2_{-\rho-i}\rangle= 
\langle h^2\rangle U_i,\hspace{2em}
\langle K^2\rangle=\langle k^2_{-\rho-\Delta-i}\rangle= 
\langle h^2\rangle V_i,\\
&&\langle H^{\prime 1}\rangle
=\langle h^{\prime 1}_{-\sigma-i}\rangle
=\langle h^{\prime 1}\rangle U'_i,\hspace{1.5em}
\langle K^{\prime 1}\rangle
=\langle k^{\prime 1}_{-\sigma-\Delta'-i}\rangle
=\langle h^{\prime 1}\rangle V'_i,\\ 
&&\langle R\rangle =\langle r_{(\rho+\sigma)/2}\rangle
=\langle r\rangle,\hspace{3.4em}
\langle\bar R \rangle=\langle \bar r_{-(\rho+\sigma)/2}\rangle
=\langle \bar r\rangle,
\end{eqnarray}
with
\begin{equation}
\langle h^2\rangle\simeq \langle h^{\prime 1}\rangle
\simeq m_{\rm SUSY},
\hspace{1.5em}
\langle r\rangle\simeq\langle\bar r\rangle\simeq
\sqrt{m_{\rm SUSY}M}.
\end{equation}
We note that we must integrate out all of the superheavy particles
before substituting these VEVs for the relevant operators.

There are some points to be mentioned for the estimation of $m_G$.
First, $G^0$ resides dominantly in $r$ and $\bar r$ with the fraction 
of the order $1$.
Second, $h^2$ and $h^{\prime 1}$ contains $G^0$ with the fraction 
on the order of $\sqrt{m_{\rm SUSY}/M}$.
Thirdly, the $A$-terms corresponding to the couplings
(\ref{PQ-br-1})$\sim$(\ref{PQ-br-8}) directly
break the $U(1)_{\rm PQ}$ symmetry.
Finally, the supersymmetric $F$-term potential breaks this 
symmetry only in the cross terms with the different $U(1)_{\rm PQ}$
charges.

Let us give the results of the analysis of $m_G$.
The couplings (\ref{PQ-br-1})$\sim$(\ref{PQ-br-8}) are classified into
the three categories:
\begin{eqnarray}
&&1.\ \ \rho+\sigma\leq 1\hspace{2em}\mbox{for any}\ \Delta, \Delta',\\
&&2.\ \ \rho+\sigma=2\hspace{2em}\mbox{for any}\ \Delta, \Delta',\\
&&3.\ \ 1<\rho+\sigma\neq 2\hspace{1em}\mbox{and}
\hspace{1em}(\rho+\sigma)/2\leq
\Delta,\Delta'.
\end{eqnarray}
The dominant couplings for each category
that give the largest $m_G$ are 
\begin{eqnarray}
&&1.\ \ R\bar R'S+\bar RR'\bar S,\\
&&2.\ \ RR\bar R'+\bar R\bar RR',\\
&&3.\ \ \bar H^{(\prime)}K^{(\prime)}R
+H^{(\prime)}\bar K^{(\prime)}\bar R.
\end{eqnarray}
The category 1 couplings give
\begin{equation}
m_G^2\sim \frac{m_0}{M^5}\langle r^6\rangle
+\frac{m_0}{M^3}\langle hh'r^2\rangle 
=O\left(\frac{m_{\rm SUSY}^4}{M^2}\right).
\end{equation}
The category 2 and 3 couplings give
\begin{equation}
m_G^2\sim \frac{m_0}{M^2}\langle hh'r\rangle
=O\left(\frac{m_{\rm SUSY}^{7/2}}{M^{3/2}}\right).
\label{category-2-3}
\end{equation}

We must also consider the possibility that the $U(1)_{\rm PQ}$
symmetry is broken in the couplings of the matter multiplets
to the superheavy multiplets that we have been discarding.
Even in this case, it is confirmed that the largest $m_G$ is
limited by (\ref{category-2-3}).

These results show that $G^0$ acquires only a tiny mass even if the 
$U(1)_{\rm PQ}$ symmetry is explicitly broken.
Although the coupling of $G^0$ to quarks and leptons are
suppressed by the factor $\sqrt{m_{\rm SUSY}/M}$,
its effects may be detectable in the future experiments.

\section{Conclusion}

First of all, we should state that our trial of solving the strong CP 
problem has not been completed not only due to the reason
that we could not give the proof of the absence of the radiative
corrections to $\theta=0$ but also due to the reason that we are 
not yet able to give the explicit form of $W[\mbox{finite dim}.]$
for $\Psi$'s, which gives the desired VEVs through their equations
of motions
\begin{equation}
\frac{\partial W[\mbox{finite dim}.]}{\partial \Psi\mbox{'s}}=0.
\label{Psi-eq-motion}
\end{equation}
The results of the present study bring much information
on the structure of $W[\mbox{finite dim}.]$.
The solution of (\ref{Psi-eq-motion}) must realize
all required VEVs satisfying the conditions (\ref{arg-1-net}),
(\ref{arg-2}), and (\ref{Option-1}) or (\ref{Option-2}). 
They clearly suggest that $\Psi$'s related to the quarks should have the
specific couplings to $\Psi$'s related to the higgses.
These results will give a valuable hint in the future study 
of determining the structure
of $W[\mbox{finite dim}.]$.
We may have a chance to understand why we have three
generations of quarks and leptons in our universe.

Second, the present model predicts, in the reasonable level of probability, 
the CKM matrix in the form presented in (\ref{CKM-II,I-1}).
Since we have not determined the matrices $O_u$ and $O_d$,
this form of the CKM matrix is not able to derive any information
from the present experimental observations.
However, when we finish the analysis of the mass hierarchies
of the quarks and the leptons, 
we will be able to determine the phase structure
of the CKM matrix.
If the result reproduces the form (\ref{CKM-II,I-1}),
the present model will obtain a reliable experimental support.

Thirdly, it should be stressed that the supersymmetry plays an
indispensable role in realizing the chiral world in the low-energy physics
from originally vectorlike theory through the spontaneous breakdown
of the P-C-T-invariance by the VEVs of the ``chiral $\Psi$'s''.
If $\Psi$'s were merely the real scalar fields, their VEVs will not be 
able to realize the chiral world.
In this context, it should be noted that the low-energy physics
must be described by the MSSM.
Although we simply introduced the superpotential (\ref{MU}) 
to induce the $\mu$-term,
we are not allowed to take the essentially different ways.
The $SU(1,1)$ symmetry rejects, for example, the Next-to-MSSM 
\cite{ref:NMSSM} 
because the coupling $S^3$ is forbidden by this symmetry.
Therefore, the experimental verification of the MSSM is crucial
for the present framework of the model.

The related problem is on the form of the K\"ahler potential $K$.
We assumed (\ref{Kaeler}) for its form.
Although this form seems to be plausible,
it is sure that, if we perform the loop
expansion for the quantum effects of the model,
we will have the additional
terms in $K$.
Obviously, we should give a definite answer to the 
question why the MSSM requires so
stringent degeneracy in the soft masses of the squarks and the sleptons
to survive in the progress
of the experimental study.
One simple way of thinking is that the present model is the 
``effective theory'' retaining only the indispensable
matter multiplets to reproduce the MSSM at the low-energy,
discarding other multiplets $Z^d$, $\bar Z^d$, and $\Psi^d$ that
the ``full theory'' contains.
It may not be unreasonable to expect the quantum effects of the
``full theory''  preserve the form of the K\"ahler potential 
(\ref{Kaeler}).
There will, however, be another possibility.
The $SU(1,1)$ gauge symmetry itself may play an essential role in this
problem within the basic framework of the model
as we observed a little bit in (\ref{K-add}).
This will be an exciting and challenging subject.

Finally, we would like to discuss the value of $c$,
which has been left undetermined.
We presume
\begin{equation}
c=-g\ (=-3).
\label{c=-g=-3}
\end{equation}
Suppose all quarks are superheavy $(g=0)$
and we have some integer $c=c(0)$:
\begin{equation}
c(g=0)=1+1+1+\cdots =c(0).
\end{equation}
When we have $g$ generations of
chiral quarks, we will lose the first $g$ terms of 1's
of this expression, and we will have
\begin{equation}
c(g)=c(0)-g.
\end{equation}
The most plausible value seems to be $c(0)=0$. In this case, all
superheavy quarks (when $g=0$) decouple from $\theta$. Remember that the
superheavy quarks must not decouple only when we have chiral quarks. In
fact, $(\ref{c=-g=-3})$ is desirable for us because we need not mention
anything on the superheavy colored multiplets with $g=0$, 
which  the ``full theory'' will contain.

We examined the strong CP problem assuming the specific form of the
superpotential for quarks and higgses. However, main ingredients of the
result of this study will not be modified even if the chiral quarks and
higgses are generated through more complicated superpotential as far as
they are generated with the definite mixing parameters $\epsilon_f$'s
with the weight 3, and $\epsilon$ and $\epsilon'$ with the weight $-1$,
which give the characteristic structure of the Yukawa coupling matrices.

\section*{Acknowledgments}

The authors would like to thank K. Harada, K. Yoshioka, K. Kojima and
H. Sawanaka for helpful discussions.
This work is supported in part by a grant-in-aid
for the scientific research on priority area ($\sharp$ 441)
``Progress in elementary particle physics of the 21st century
through discoveries of Higgs boson and supersymmetry''
(No. 16081209) from the Ministry of Education, Culture,
Sports, Science and Technology of Japan.

\end{document}